\pdfoutput=1
\documentclass[fleqn,usenatbib]{mnras}

\usepackage{newtxtext,newtxmath}
\usepackage{amsmath}
\usepackage{graphics,graphicx}
\usepackage{natbib}
\usepackage{setspace}
\usepackage{caption,setspace}
\newcommand{\angstrom}{\mbox{\normalfont\AA}}
\usepackage{bm}

\usepackage[dvipsnames]{xcolor}

\usepackage{mathtools}
\usepackage{enumitem}
\usepackage{lipsum}
\setlist[itemize]{leftmargin=*}

\usepackage{floatrow}
\usepackage[outercaption]{sidecap}    
\sidecaptionvpos{figure}{t}

\usepackage[T1]{fontenc}
\usepackage{ae,aecompl}
\hypersetup{pdfauthor={R. K. Cochrane},
               pdftitle={Resolving a dusty, star-forming SHiZELS galaxy at z=2.2 with HST, ALMA and SINFONI on kiloparsec scales},
               pdfkeywords={galaxy evolution, high redshift galaxies, submillimetre galaxies},
               bookmarksnumbered=true}
               
\title[SHiZELS-14: Resolving a dusty, star-forming galaxy at $z=2.2$]{Resolving a dusty, star-forming SHiZELS galaxy at $\mathbf{z=2.2}$ with \textbf{\emph{HST}}, ALMA and SINFONI on kiloparsec scales}

\author[R.K. Cochrane et al.]{R. K. Cochrane,$^{1,2}$\thanks{E-mail: rachel.cochrane@cfa.harvard.edu}
P. N. Best,$^{2}$
I. Smail, $^{3}$
E. Ibar, $^{4}$
C. Cheng, $^{5,6,4}$
A. M. Swinbank,$^{3}$\newauthor
J. Molina, $^{7}$ 
D. Sobral, $^{8}$
U. Dudzevi\v{c}i\={u}t\.{e}$^{3}$
\vspace{0.1cm}\\
$^{1}$Harvard-Smithsonian Center for Astrophysics, 60 Garden St. Cambridge, MA 02138, USA\\
$^{2}$SUPA, Institute for Astronomy, Royal Observatory Edinburgh, EH9 3HJ, UK\\
$^{3}$Centre for Extragalactic Astronomy, Department of Physics, Durham University, South Road, Durham DH1 3LE, UK  \\
$^{4}$Instituto de F\'{i}sica y Astronom\'{i}a, Universidad de Valpara\'{i}so, Avda. Gran Breta\~{n}a 1111, Valpara\'{i}so, Chile \\
$^{5}$Chinese Academy of Sciences South America Center for Astronomy, National Astronomical Observatories, CAS, Beijing 100101, China\\
$^{6}$CAS Key Laboratory of Optical Astronomy, National Astronomical Observatories, Chinese Academy of Sciences, Beijing 100101, China\\
$^{7}$Kavli Institute for Astronomy and Astrophysics, Peking University, 5 Yiheyuan Road, Haidian District, Beijing 100871, P.R. China\\
$^{8}$Department of Physics, Lancaster University, Lancaster, LA1 4YB}

\date{Accepted 2021 February 10. Received 2021 February 10; in original form 2020 August 21}

\pubyear{2021}

\volume{{\rm in press}}
   
\begin{document}
\label{firstpage}
\pagerange{\pageref{firstpage}--\pageref{lastpage}}
\maketitle

\begin{abstract}
We present $\sim0.15''$ spatial resolution imaging of SHiZELS-14, a massive ($M_{*}\sim10^{11}\,\rm{M_{\odot}}$), dusty, star-forming galaxy at $z=2.24$. Our rest-frame $\sim1\rm{kpc}$-scale, matched-resolution data comprise four different widely used tracers of star formation: the $\rm{H}\alpha$ emission line (from SINFONI/VLT), rest-frame UV continuum (from {\it{HST}} F606W imaging), the rest-frame far-infrared (from ALMA), and the radio continuum (from JVLA). Although originally identified by its modest $\rm{H}\alpha$ emission line flux, SHiZELS-14 appears to be a vigorously star-forming ($\rm{SFR}\sim1000\,\rm{M_{\odot}yr^{-1}}$) example of a submillimeter galaxy, probably undergoing a merger. SHiZELS-14 displays a compact, dusty central starburst, as well as extended emission in $\rm{H}\alpha$ and the rest-frame optical and FIR. The UV emission is spatially offset from the peak of the dust continuum emission, and appears to trace holes in the dust distribution. We find that the dust attenuation varies across the spatial extent of the galaxy, reaching a peak of at least $A_{\rm{H}\alpha}\sim5$ in the most dusty regions, although the extinction in the central starburst is likely to be much higher. Global star-formation rates inferred using standard calibrations for the different tracers vary from $\sim10-1000\,\rm{M_{\odot}yr^{-1}}$, and are particularly discrepant in the galaxy's dusty centre. This galaxy highlights the biased view of the evolution of star-forming galaxies provided by shorter wavelength data.
\end{abstract}
\section{Introduction}
Galaxy surveys have long shown that star-formation rates within individual galaxies increase towards high redshift. At a given stellar mass, typical star-formation rates increase by over an order of magnitude between the present day and the peak of cosmic star formation at $z\sim2$ \citep{Sobral2012,Madau2014,Speagle2014}. This is thought to reflect the large reservoirs of molecular gas that cool from the high rates of gas accretion onto galaxies' host halos in the early universe \citep{Tacconi2010,Tacconi2013,Tacconi2017,Papovich2016,Falgarone2017a,Jimenez-Andrade2018,Dudzeviciute2019}. \\
\indent Although highly luminous dusty galaxies are rare at $z=0$ and known as `ultra luminous infrared galaxies' (ULIRGs, with total infrared luminosities $L_{\rm{TIR}}>10^{12-13}\,\rm{L_{\odot}}$), galaxies with typical ULIRG luminosities are more common around the peak of cosmic star formation \citep{Smail1997,Barger1998}. Submillimeter galaxies (SMGs; \citealt{Blain2002}) are ULIRGs at high redshift with bright submillimeter fluxes that suggest star-formation rates (SFRs) of $\sim100-1000\,\rm{M_{\odot}yr^{-1}}$. Sustained star-formation rates of this magnitude have the potential to form massive galaxies (with stellar masses of $\sim10^{11}\,\rm{M_{\odot}}$) on sub-Gyr timescales \citep{Simpson2014,Dudzeviciute2019}. \cite{Chapman2005} found that the volume density of SMGs increases by a factor of $\sim1000$ between $z=0$ and $z=2.5$, with the redshift distribution peaking at $z\sim2.0-2.5$ (see also \citealt{Koprowski2014,Simpson2014,Danielson2017,Stach2019}; recent studies using larger samples derive a redshift distribution that peaks slightly higher). SMGs at $1<z<5$ appear to account for $\sim20-30$ per cent of the total comoving star-formation rate density at these redshifts \citep{Swinbank2014,Smith2017a,Dudzeviciute2019}. Even in less far-infrared (FIR)-luminous high redshift galaxies, a significant amount of star formation is obscured by dust. \cite{Dunlop2017} combined long- and short-wavelength data from two premier observatories: the Atacama Large Millimeter Array (ALMA, probing the dust continuum emission at 1.3mm) and the {\it{Hubble Space Telescope}} ({\it{HST}}, Wide Field Camera 3, probing rest-frame UV), in the well-studied Hubble Ultra Deep Field \citep[e.g.][]{Bouwens2010,Oesch2010,Illingworth2013,Dunlop2013}. These complementary data enabled them to confirm that $\sim85$ per cent of the total star formation at $z\sim1-3$ is enshrouded in dust. Emission from the most massive galaxies is most highly attenuated: for galaxies with $M_{*}\sim5\times10^{10}\,\rm{M_{\odot}}$, they suggest a ratio of obscured to unobscured star formation of $\sim50$. However, lower mass galaxies are still affected, with the ratio decreasing to $\sim5$ for galaxies with $M_{*}\sim5\times10^{9}\,\rm{M_{\odot}}$ \citep[see also][]{Magnelli2020}. \\
\indent While studies of wide areas are important in tracking the evolving properties of star-forming galaxies and the build-up of stellar mass in the Universe, understanding the physical processes of star formation within individual galaxies requires higher angular resolution. Until recently, resolved studies of distant star-forming (SF) galaxies tended to be based on observations from near-infrared integral field unit spectrographs, which probe rest-frame optical emission lines such as $\rm{H}\alpha$ and [O{\small{III}}] at $z\sim2$ \citep[e.g.][]{Genzel2008,Swinbank2012a,Reddy2015,Stott2016,Simons2017}, or from {\it{HST}} at rest-frame UV wavelengths \citep[e.g.][]{Wuyts2012,Fisher2017a}. These claim a physical picture in which star formation takes place within massive clumps embedded in turbulent disk structures \citep[][]{Genzel2008,Elmegreen2013,Genzel2013,Guo2015,Guo2017b,Soto2017}. Emission at these short wavelengths is, however, strongly attenuated by dust, and the significant global obscuration of star formation at $z<4$ suggests that our understanding of galaxy evolution from short-wavelength studies is likely to be highly biased by dust, even at high spatial resolution. As such, the importance and even the reality of these clumps has been questioned \citep[e.g.][]{Hodge2016,Hodge2019,Ivison2020}. Indeed, star formation in the dustiest regions of high redshift galaxies is expected to be totally hidden from view \citep{Simpson2016}.\\
\indent Recent work made possible by new submillimeter interferometers, in particular ALMA, which offers both high sensitivity and spatial resolution, has focused on characterising the spatially-resolved properties of high redshift galaxies at long wavelengths (see the recent review by \citealt{Hodge2020}). The spatial extent of dust emission and molecular gas has been of particular interest in recent years. The dust continuum emission and CO emission appear very compact for distant ($z>1$), sub-millimeter-bright galaxies, with typical effective radii $\sim1-2\,\rm{kpc}$ \citep{Simpson2015a,Tadaki2016,Tadaki2017,Tadaki2018,Hodge2016,Hodge2019,Oteo2017,Strandet2017,Rivera2018,Lang2019,Gullberg2019,Dudzeviciute2019}. A number of studies have shown that these sizes are comparable to the optical sizes of $z\sim1-2$ compact quiescent ellipticals, galaxies that must have formed a huge amount of stellar mass and then quenched early \citep{Krogager2014,Onodera2015,Belli2016,Lang2019}. This, together with the large estimated stellar masses of SMGs ($M_{*}\sim10^{11}\rm{M_{\odot}}$; \citealt{Hodge2019,Dudzeviciute2019}) has fuelled speculation that the SMGs detected at $z\sim3-6$ are the progenitors of $z=2$ massive ellipticals \citep[e.g.][]{Toft2014,Simpson2014,Oteo2017,Gomez-Guijarro2018,Tadaki2020a}, possibly tracing a rapid phase of bulge-building \citep[e.g.][]{Tadaki2016,Simpson2016,Nelson2018}. \\
\indent However, observations of compact dust continuum sizes are in contrast to the extended, clumpy structures traced by {\it{HST}} imaging \citep{Chen2015,Barro2016,Hodge2015,Hodge2016,Hodge2019,Rujopakarn2019}. In some cases, kpc-scale offsets have been found between the peaks of the FIR and UV emission \citep{Hodge2015,Tadaki2016,Chen2017,Rivera2018}. These offsets could potentially bias interpretations of global measurements (particularly for fits to photometry that focus solely on the rest-frame optical to near-infrared, but also for `energy-balance' spectral energy distribution fitting). Indeed, \cite{Simpson2016} argue that attenuation in the dusty regions of SMGs is so great that essentially all the co-located stellar emission is obscured at optical-to-near-infrared wavelengths; for $\sim30$ per cent of their sample, the data available at these wavelengths is insufficient to put constraints on photometric redshifts and stellar masses (see also work on `NIR-dark' sources; e.g. \citealt{Simpson2014,Franco2018,Wang2019a,Dudzeviciute2019,Smail2020})  \\
\indent Overall, it has become clear that drawing conclusions from single-wavelength surveys, especially in the rest-frame UV, is subject to substantial bias and uncertainty, even where data is at high angular resolution. In this paper, we present multi-wavelength, $0.15''$-resolution imaging of SHiZELS-14, a highly star-forming, $\rm{H}\alpha$-selected galaxy at $z=2.24$. Of the ALMA-studied SHiZELS parent sample (which is presented in a companion paper, \citealt{Cheng2020}), SHiZELS-14 is the most FIR luminous, with the largest of all $\rm{H}\alpha$-derived effective radii ($4.6\pm0.4\,\rm{kpc}$)
\citep{Swinbank2012a,Swinbank2012b,Gillman2019}. Although its $\rm{H}\alpha$ flux is modest, it displays SMG-like dust continuum emission. Our observations comprise matched-resolution imaging of the $\rm{H}\alpha$ emission line (from SINFONI/VLT), rest-frame UV and optical continuum (from {\it{HST}}), and the rest-frame far-infrared (from ALMA), as well as the radio continuum (from the Karl G. Jansky Very Large Array; JVLA). We find bright, extended structures in the multi-wavelength imaging, with clear clumps in $\rm{H}\alpha$ and extended dust continuum emission. Given this extended structure and the high signal-to-noise that results from its high SFR, we have been able to resolve star formation on kpc scales at multiple wavelengths.\\
\indent The structure of this paper is as follows. In Section \ref{sec:overview}, we provide an overview of the data available for our study of SHiZELS-14. We review the high quality, but less well-resolved multi-wavelength data available from imaging of the COSMOS field, and present the new $0.15''$ resolution imaging from SINFONI/VLT, {\it{HST}}, ALMA and JVLA. We discuss the astrometric alignment of these data in Section \ref{sec:astrometry}. In Section \ref{sec:global}, we present the global properties of SHiZELS-14 that may be inferred from spectral energy distribution (SED) fitting. In Section \ref{sec:resolved_comparison}, we present maps of the spatially-resolved SFRs inferred from different SFR indicators, and derive a spatially-resolved dust attenuation map. 
In Section \ref{sec:smg_comparison}, we compare the properties of SHiZELS-14 to the submillimeter galaxy population. In Section \ref{sec:conclusions}, we summarise our results.\\
\indent We assume a $\Lambda$CDM cosmology with $H_{0} = 70\,\rm{km}\,\rm{s}^{-1}\,\rm{Mpc}^{-1}$, $\Omega_{M} = 0.3$ and $\Omega_{\Lambda} = 0.7$. We use a \cite{Kroupa2002} initial mass function (IMF).
 
\section{Observations and data reduction}\label{sec:overview}
The High-Redshift(Z) Emission Line Survey, HiZELS, used a combination of narrow-band and broad-band filters to select star-forming galaxies via their emission line fluxes \citep{Sobral2012,Sobral2015} in fields with high-quality multi-wavelength coverage (COSMOS, UDS \& SA22). This survey has yielded thousands of $\rm{H}\alpha$ emitters at $z=0.4$, $0.8$, $1.47$ \& $2.23$, providing sufficiently large samples to constrain $\rm{H}\alpha$ luminosity functions, stellar mass functions and halo environments of typical star-forming galaxies around the peak of cosmic star formation \citep{Geach2008,Sobral2009,Sobral2010,Sobral2014,Cochrane2017,Cochrane2017a}.\\
\indent As well as providing the sample sizes for population studies such as these, HiZELS has also provided parent samples for more detailed follow-up observations \citep{Sobral2013,Magdis2016,Stott2016,Molina2017,Molina2019,Gillman2019}. In particular, by exploiting the wide area HiZELS coverage, a sample of bright $\rm{H}\alpha$ emitters ($f_{\rm{H}\alpha} > 0.7\times 10^{-16}\,\rm{erg}\,\rm{s}^{-1}\rm{cm}^{-2}$) which by chance lie within $30''$ of bright natural guide stars ($R$<15) could be identified and targeted for IFU spectroscopy of the $\rm{H}\alpha$ line using adaptive optics with the SINFONI Integral Field Unit on the Very Large Telescope (VLT). This campaign, known as SINFONI-HiZELS (SHiZELS), yielded high-resolution spectral maps for $20$ galaxies at $z=0.8$, $z=1.47$ and $z=2.23$ at $\sim0.15''$ (rest-frame $\sim1\,\rm{kpc}$) resolution \citep[see][]{Swinbank2012a,Swinbank2012b,Molina2017,Gillman2019}.\\
\indent We complemented these data with imaging at similar angular resolution but different wavelengths. Nine HiZELS galaxies were targeted at $\sim0.2''$ resolution with ALMA (Band 6 or 7, depending on redshift), to map the dust continuum emission (see \citealt{Cheng2020}). UVIS Imaging in the rest-frame UV (F606W) and rest-frame optical (F140W) filters obtained during {\it{HST}} Cycle 24 completes this dataset. We now have FIR-UV-$\rm{H}\alpha$ matched-resolution observations of a small sample of HiZELS galaxies. Since these galaxies are $\rm{H}\alpha$-selected, they are likely to be a less biased sub-sample of the high-redshift star-forming galaxy population than UV-selected samples, which target the bluest and least dusty galaxies, at an epoch where dust is important \citep[see][]{Oteo2015}. \\
\indent Here, we present data for SHiZELS-14, which is the brightest, most extended and more extreme source in our sample. SHiZELS-14 (10:00:51:6 +02:33:34.5) is a $z=2.24$ galaxy, with high stellar mass ($M_{*}\sim10^{11}\,\rm{M_{\odot}}$; \citealt{Swinbank2012a,Laigle2016}), and a star-formation rate of $\sim1000\,\rm{M_{\odot}yr^{-1}}$ These properties enable a detailed investigation of the multi-wavelength extended structures of this galaxy. In the following subsections, we provide details of the new high-resolution imaging we have recently obtained as part of the SHiZELS campaign. We present new radio continuum imaging from the JVLA (at comparable angular resolution to the other new imaging), which were obtained only for SHiZELS-14. We also describe the existing data available for our multi-wavelength characterisation of this galaxy.

\begin{figure*}
\includegraphics[width=\columnwidth]{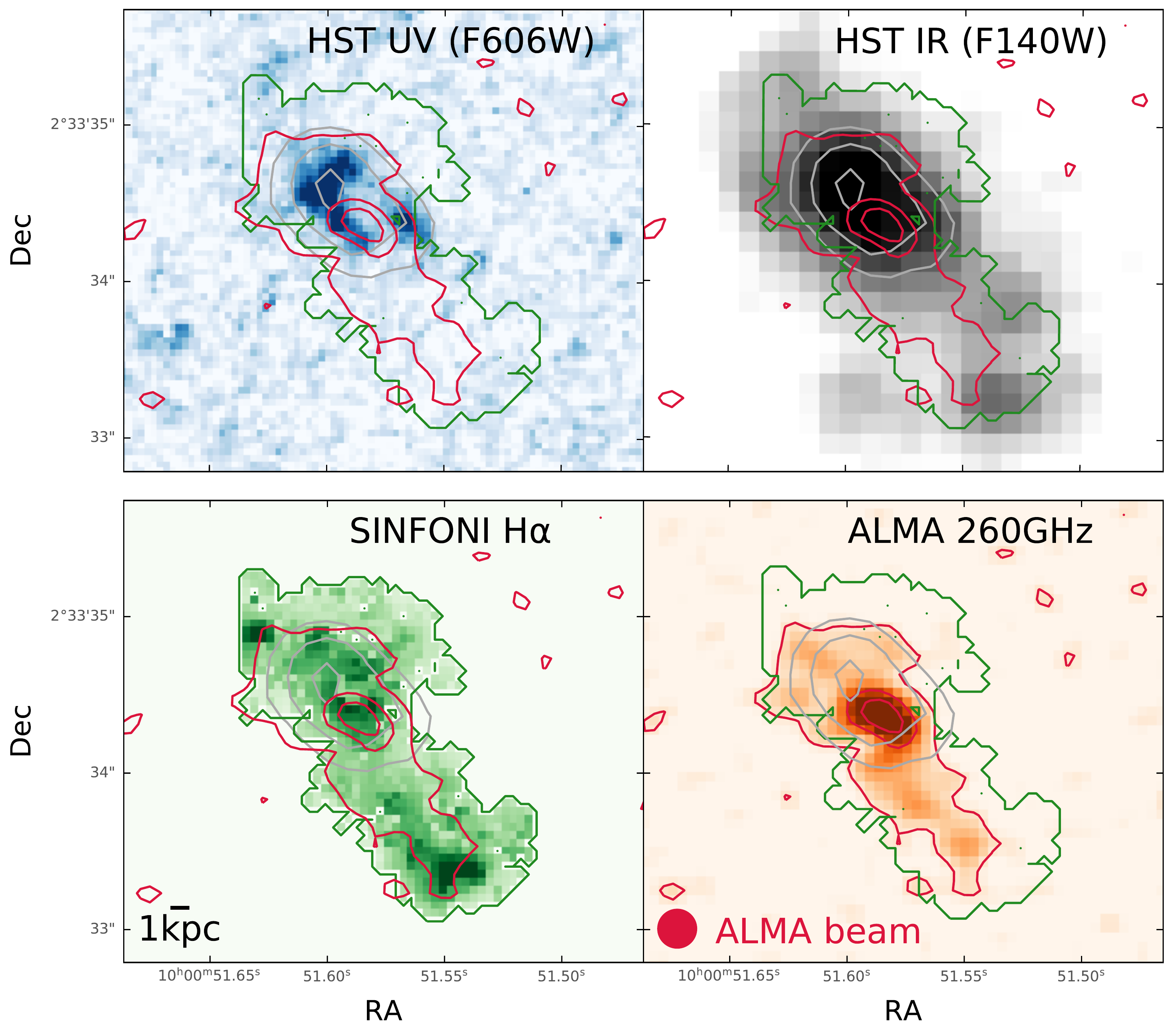}
\caption{Astrometrically-calibrated, high-resolution observations of SHiZELS-14 in the rest-frame UV ({\it{HST}} F606W filter; upper left panel), rest-frame optical ({\it{HST}} F140W filter; upper right panel), $\rm{H}\alpha$ (SINFONI/VLT; lower left) and dust continuum (rest-frame $370\,\micron$ imaging from ALMA, reduced with Briggs weighting; lower right). The red contours on all panels outline the ALMA dust continuum emission at 50, 200, and 300$\,\mu\rm{Jy\,beam^{-1}}$. The green contours outline the $3\sigma$ emission $\rm{H}\alpha$ emission from SINFONI as described in Section \ref{sec:sinfoni}. Pale grey contours outline the peak of the F140W image. The emission imaged by SINFONI, ALMA and the {\it{HST}} F140W filter span the same extended region, but display very different morphologies. The peaks of the short-wavelength emission are clearly offset from the peaks of the dust continuum emission. This is particularly striking for the F606W UV emission, which is concentrated in regions with little dust emission and does not extend down to the southern regions that are clearly probed by the other bands.}
\label{fig:four_images}
\end{figure*}

\begin{figure*}
\includegraphics[width=\columnwidth]{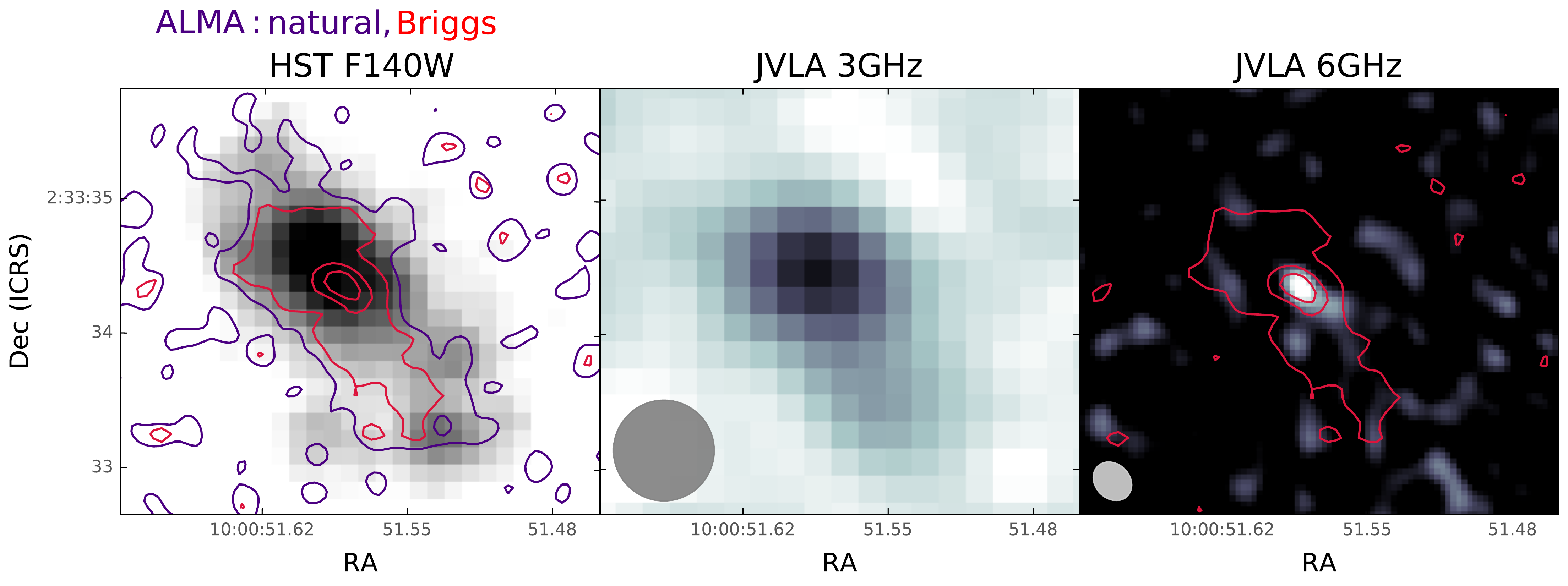}
\caption[]{Left: the {\it{HST}} F140W image, with contours of the $260\,\rm{GHz}$ ALMA data with two weightings overlaid. The image produced using natural weighting is shown with purple contours tracing $25\,\mu\rm{Jy\,beam^{-1}}$. The red contours outline the Briggs-weighted image ($50$, $200$, and $300\,\mu\rm{Jy\,beam^{-1}}$, as in Figure \ref{fig:four_images}). The slightly lower angular resolution natural-weighted image shows flux towards the North East and the South West, in the regions with extended F140W flux. This gives us confidence in the astrometric alignment of the images. Centre: $0.75''$ imaging from the VLA-COSMOS $3\,\rm{GHz}$ Large Project \citep{Smolcic2017}. Right: new, $\sim0.33''$, $4-8\,\rm{GHz}$ continuum imaging from the JVLA. The peak of the radio continuum emission coincides with the peak of the ALMA map. On both radio images, the beam is plotted in grey.}
\label{fig:alma_reductions}
\end{figure*}

\subsection{Resolved $\rm{H}\alpha$ emission from SINFONI}\label{sec:sinfoni}
SINFONI observations of SHiZELS-14 took place in March 2010, in good seeing and photometric conditions ($\sim0.6''$), with total exposure time $12\rm{ks}$ (each individual exposure was $600\rm{s}$). This yielded the sub-kpc resolution $\rm{H}\alpha$ map shown in the lower left-hand panel of Figure \ref{fig:four_images}. SHiZELS-14 was the only $z\sim2.2$ source resolved in this initial \cite{Swinbank2012a,Swinbank2012b} campaign (though note that five more $z\sim2.2$ galaxies were targeted in the campaign presented by \citealt{Molina2017}). \\
\indent Data reduction and analysis procedures are outlined in full in \cite{Swinbank2012a,Swinbank2012b} (see also \citealt{Molina2017} and \citealt{Gillman2019}). In summary, the SINFONI {\small{ESOREX}} data reduction pipeline was used to perform extraction, flat fielding and wavelength calibration, and to create the data cube for each exposure. These data cubes were then stacked and combined using an average with a $3\sigma$ clip, to reject cosmic rays and sky line residuals. Flux calibration was performed using observations of standard stars taken immediately before/after science exposures, which were reduced in the same way. $\rm{H}\alpha$ and [N{\sc ii}]$\lambda \lambda6548,6583$ emission lines were fitted on a pixel-by-pixel basis, using a $\chi^{2}$ minimisation procedure. This yielded intensity, velocity, and velocity dispersion maps. An angular resolution of $\sim0.15''$ was achieved. The spectral resolution of the instrument is $\lambda/\Delta \lambda \sim4500$. \\
\indent The $\rm{H}\alpha$ flux derived from the SINFONI observations of SHiZELS-14 was $1.6\pm0.1\times10^{-16}\,\rm{erg\,s^{-1}\,cm^{-2}}$. The $\rm{H}\alpha$-derived effective radius is $4.6\pm0.4\,\rm{kpc}$ \citep{Swinbank2012b}. Using the same SINFONI data, \cite{Gillman2019} derive $V_{\rm{rot}}/\sigma=0.6\pm0.3$ (indicating that SHiZELS-14 is dispersion dominated), though, as noted by \citep{Swinbank2012a}, this galaxy shows a substantial (in their paper, $480\pm40\,\rm{km\,s^{-1}}$) peak-to-peak velocity gradient. \cite{Swinbank2012a} comment that the one- and two-dimensional velocity fields are consistent with an early-stage prograde encounter. This suggests that the disordered morphology and extreme star formation may be related to a merger event.

\subsection{Resolved UV and optical light from \textbf{\emph{HST}}}
SHiZELS-14 was observed over two  {\it{HST}} orbits during Cycle 24 (Program 14719, PI: Best). One orbit ($2700\,\rm{s}$ exposure) used the WFC3/UVIS F606W filter, and the other used the WFC3/IR F140W filter. Orbits were split into a 3-point dither pattern in the UVIS channel, as a compromise between maximising sensitivity and sub-sampling the point spread function (PSF). Since angular resolution was preferred over sensitivity in the IR channel, a 4-point dither pattern was used for these orbits. At $z=2.24$ (the redshift of SHiZELS-14), the filters correspond to the rest-frame near-UV at $\sim1900\,\angstrom$, and the rest-frame optical at $\sim4400\,\angstrom$. Our observations were designed to span the $4000\,\angstrom$ break, and therefore sample both young and more evolved stellar populations, in line-free regions of the galaxy spectrum. The {\it{HST}} images are made using standard {\it{HST}} procedures and shown in the upper panels of Figure \ref{fig:four_images}. We derive the effective radius of the F140W image via a two-dimensional S\'{e}rsic profile fit, obtaining effective radius along the semi-major axis $R_{e,\,\rm{opt}}^{\rm{maj}} = 4.6\pm0.2\,\rm{kpc}$ (in good agreement with the $\rm{H}\alpha$ measurement) and axial ratio $q=0.64$ (with a S\'{e}rsic index fixed at $n=1$; fitting this parameter gives $n=0.9$).

\subsection{Resolved far-infrared emission from ALMA}\label{sec:alma_data}
SHiZELS-14 was observed with ALMA during August 2016 as part of ALMA Cycle 3 (project code 2015.1.00026.S, PI: Ibar). Our observations, taken in configuration C36-5, used Band 6 ($260\,\rm{GHz}$, $7.5\,\rm{GHz}$ bandwidth). The time on-source was $26\,\rm{minutes}$. We used flux calibrator J1058+0133 and phase calibrator J0948+0022. These observations resolved the rest-frame $840\,\rm{GHz}$ ($367\micron$) emission of SHiZELS-14 at $\sim0.2''$ resolution. \\
\indent The image was manually cleaned down to $3\sigma$ ($\rm{rms}\sim25\,\mu\rm{Jy\,beam^{-1}}$) at the source position. We used Briggs (robust=0.5) visibility weighting, which assigns higher weights to longer baselines, producing an image with higher angular resolution (see the image in the lower right-hand panel of Figure \ref{fig:four_images}). To investigate the impact of visibility weighting on the reduced ALMA image, we re-imaged the ALMA data using a natural weighting, which weights visibilities only by the rms noise (see the left-hand panel of Figure \ref{fig:alma_reductions}). This method minimises the noise level but provides poorer angular resolution, given that the density of visibilities falls towards the outskirts of the $uv$-plane and there is thus higher noise in the longer-baseline visibilities. Using the re-reduced, lower angular resolution natural-weighted image, we probe to slightly lower flux density per beam. This will be used to assess the quality of our astrometric calibration in Section \ref{sec:astrometry}.\\
\indent SHiZELS-14 has an observed-frame $260\,\rm{GHz}$ flux density of $2.7\pm0.2\,\rm{mJy}$. It displays a compact, $\sim3\,\rm{kpc}$ diameter core of dust emission, with extended emission contributing substantially to the flux. Its effective radius is notably larger, due to this extended faint emission. We derive this radius using multiple methods. First, we fit a Gaussian model with varying axis ratio in the $uv$-plane, using {\small{CASA}}'s {\it{uvmodelfit}} task (see Figure \ref{fig:uv_ampl_plot}). The effective radius along the semi-major axis, $R_{e}^{\rm{maj}}$, is $4.5\pm0.2\,\rm{kpc}$, with fitted axial ratio $q=0.36\pm0.01$. A two-dimensional S\'{e}rsic profile fit in the image plane yields $R_{e,\rm{FIR}}^{\rm{maj}} = 4.6\pm0.2\,\rm{kpc}$ and $q=0.47$ (with the S\'{e}rsic index fixed at $n=1$; allowing this to vary gives $n=1.1$). These measurements of effective radius are broadly consistent with those derived from the SINFONI $\rm{H}\alpha$ and the HST rest-frame optical data.

\subsection{Existing radio observations from COSMOS-VLA}
We make use of the deep existing radio observations in the COSMOS field from the VLA-COSMOS surveys. The VLA-COSMOS Large Project \citep{Schinnerer2007} surveyed $2$ square degrees in VLA A- and C-array configurations at $1.4\,\rm{GHz}$ ($20\,\rm{cm}$). The project yielded images with rms noise $\sim10-15\,\mu\rm{Jy\,beam^{-1}}$ at angular resolution $1.5''\times1.4''$. The VLA-COSMOS Deep project \citep{Schinnerer2010} added further A-array observations at $1.4\,\rm{GHz}$ in the central region of the COSMOS field. The VLA-COSMOS $3\,\rm{GHz}$ Large Project \citep{Smolcic2017} subsequently surveyed $2.6\,\rm{deg}^{2}$ at a wavelength of $10\,\rm{cm}$ with the upgraded JVLA in A configuration, reaching a mean rms depth of $\sim2.3\,\mu\rm{Jy\,beam^{-1}}$ at $0.75''$ angular resolution. \\
\indent SHiZELS-14 is one of the sources detected by these VLA surveys. The measured flux densities are $S_{1.4\rm{GHz}} = 122\pm13\,\mu\rm{Jy}$ and $S_{3\rm{GHz}} = 68\pm4\,\mu\rm{Jy}$. From these two flux densities, we derive a radio spectral index of $\alpha=-0.77\pm0.16$ (where $S_{\nu} \propto \nu^{\alpha}$), in good agreement with measurements of star-forming galaxies \citep{Condon1992}. The lower angular resolution of the radio images limits our ability to probe resolved structure (see Figure \ref{fig:alma_reductions}, centre panel), but the source is still extended at $0.75''$ resolution. We will use the total flux density to estimate a star-formation rate later in the paper.

\subsection{New resolved radio observations from the JVLA}
We obtained C-band ($4-8\,\rm{GHz}$) observations of SHiZELS-14 using 27 antennas of the JVLA, arranged in A-array configuration ($\sim0.33''$ spatial resolution). Observations took place during October 2019, as part of Cycle 19A (project 19A-205). We used 3C147 for flux calibration, and J1024-008 for phase calibration. The data presented were obtained during a $4\,\rm{hr}$ observing block, with around $3\,\rm{hrs}$ of on-source time. \\
\indent We reduced the data using standard {\small{CASA}} calibration pipelines, and manually cleaned the images down to $2\,\mu\rm{Jy\,beam^{-1}}$. We present our Briggs-weighted image in Figure \ref{fig:alma_reductions}. We obtain a total continuum flux density of $20\pm2\,\mu\rm{Jy}$ at $6\,\rm{GHz}$. This is only roughly half the expected flux, based on the $1.4\,\rm{GHz}$ and $3\,\rm{GHz}$ data (assuming the spectral index calculated from these observations, $\alpha=-0.77$). One explanation for this is that the image is not sufficiently deep to resolve the lower surface brightness emission in the faint outskirts of the galaxy seen by ALMA. This would suggest that the radio emission takes the form of a bright compact core, with fainter extended structure (i.e. the radio emission is not driven primarily by a point source). This is consistent with the  nature of the JVLA $3\,\rm{GHz}$ image, which is clearly extended even at lower angular resolution (Figure \ref{fig:alma_reductions}, centre panel). The peak of the radio continuum emission coincides with the peak of the dust continuum emission, which gives confidence in the astrometric accuracy of our ALMA data (see Section \ref{sec:astrometry}).

\begin{figure*}
\centering
\includegraphics[width=\columnwidth]{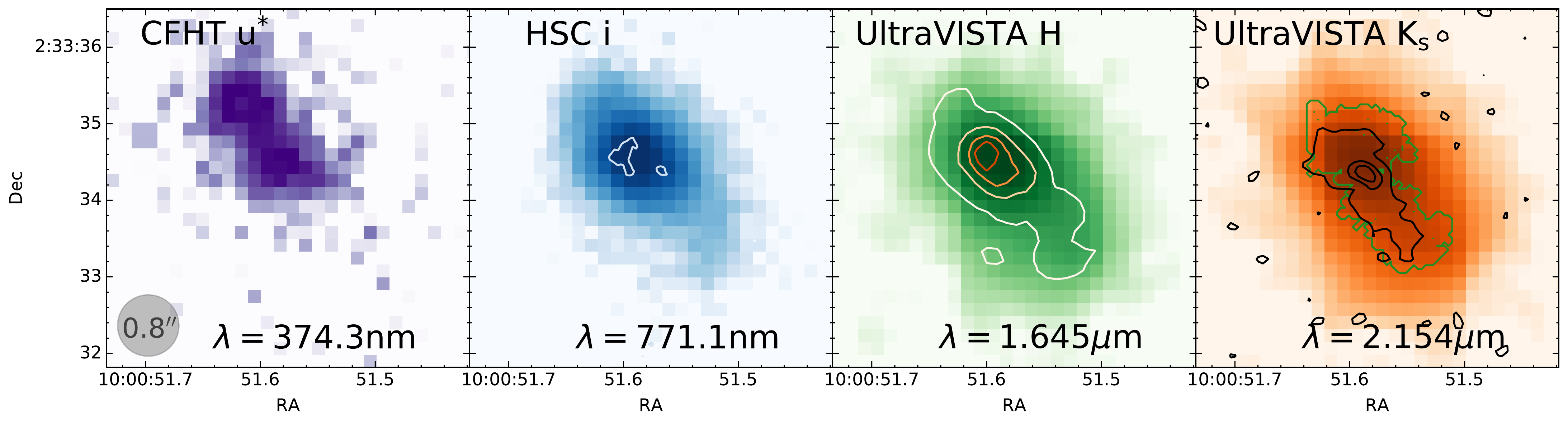}
\caption[NUV-NIR imaging of SHiZELS-14 from CFHT]{NUV-NIR imaging of SHiZELS-14 from CFHT (u* from the CLAUDS survey; \citealt{Sawicki2019}), Subaru (HSC-DR2; \citealt{Aihara2019}), and VISTA (UltraVISTA DR4; \citealt{Mccracken2015}). These observations are seeing-limited, with angular resolution $\sim0.65-0.9''$. We show the typical angular resolution on the CFHT $u^{*}$ image. We overlay contours from our resolved imaging on relevant panels. Overplotted on the HSC $i$-band image are contours from {\it{HST}} F606W imaging (blue). The contours on the UltraVISTA $H$-band image are from {\it{HST}} F140W imaging (orange). Both SINFONI $\rm{H}\alpha$ (green) and ALMA dust continuum emission (black) contours are overplotted on the UltraVISTA $K_{s}$-band image. }
\label{fig:cfht_mosaics}
\end{figure*}

\subsection{Optical/IR data from the COSMOS field}\label{sec:data1}
A wealth of lower resolution data exists for this galaxy due to its location within the well-imaged COSMOS field \citep{Scoville2007}. At NUV-optical wavelengths, COSMOS was imaged in the $u*$-band from the Canada-France-Hawaii Telescope (CFHT/MegaCam), and in six broad bands ($B$, $V$, $g$, $r$, $i$,
$z+$), 12 medium bands (IA427, IA464, IA484, IA505, IA527, IA574, IA624, IA679, IA709, IA738, IA767, and IA827), and two narrow bands (NB711, NB816), all from the COSMOS-20 survey (Subaru Suprime-Cam; \citealt{Taniguchi2007,Taniguchi2015}). $Y$-band imaging was obtained with Hyper-Suprime-Cam on Subaru (HSC; Miyazaki
et al. 2012). In the NIR, $Y$, $J$, $H$, \& $K_{s}$ data are provided by the UltraVISTA-DR2 release \citep{Mccracken2015}, which uses the VIRCAM instrument on the VISTA telescope. These are supplemented by $H$ and $K$ WIRCAM data from CFHT (McCracken et al. 2010). Mid-IR data are drawn from IRAC channels 1, 2, 3 and 4 ($3.6\,\mu\rm{m}$, $4.5\,\mu\rm{m}$, $5.8\,\mu\rm{m}$ and $8.0\,\mu\rm{m}$), collected by the {\it{Spitzer}} Large Area Survey with HSC (SPLASH survey; \citealt{Lin2017}; Capak et al. in prep). \cite{Laigle2016} collate these observations and provide an NIR-selected photometric redshift catalogue. For consistency, we use their $3''$ diameter aperture fluxes extracted for SHiZELS-14. We tabulate these measurements in Table \ref{Table:hizels_laigle} of the Appendix, along with the new measurements from this paper.\\
\indent In Figure \ref{fig:cfht_mosaics} we show deeper imaging from more recent surveys: the $u^{*}$ band (from the CLAUDS survey on CFHT; \citealt{Sawicki2019}), $i$-band (from HSC-DR2; \citealt{Aihara2019}), and in the $H$ and $\rm{K}_{s}$ bands from UltraVISTA-DR4. These show interesting differences, with emission in the $u^{*}$-band peaking to the North East compared to the $K_{s}$-band emission (this is not driven by astrometric offsets in the $u^{*}$-band data). 
 
\subsection{Data at mid-IR and far-IR wavelengths}\label{sec:data2}
We draw data at mid-IR and far-IR wavelengths from {\it{Spitzer}} and {\it{Herschel}} imaging. We adopt the $24\,\micron$ flux density from the {\it{Spitzer}} Multiband Imaging Photometer (MIPS; \citealt{Rieke2004}). The {\it{Herschel}} Multi-tiered Extragalactic Survey (HerMES; \citealt{Oliver2012}) targeted COSMOS at $100-500\,\micron$. The survey used {\it{Herschel}}-Spectral and Photometric Imaging Receiver (SPIRE) at $250\,\micron$, $350\,\micron$ and $500\,\micron$  and the {\it{Herschel}}-Photodetector Array Camera and Spectrometer (PACS) at $100\,\micron$ and $160\,\micron$. One of the main aims of the {\it{Herschel}} Extragalactic Legacy Project (HELP)\footnote{http://herschel.sussex.ac.uk} was to develop the advanced statistical tools needed to de-blend the low-resolution data from {\it{Herschel}}, in order to assign fluxes to components \citep{Hurley2017,Pearson2017}. We use these publicly available, catalogued flux densities for SHiZELS-14 (see Table \ref{Table:hizels_laigle}).\\
\indent We also use the ALMA Band 7 flux density measured by \cite{Scoville2014}. The observed-frame continuum $350\,\rm{GHz}$ total flux density is $4.7\pm0.8\,\rm{mJy}$, and the peak flux density is $1.9\pm0.3\,\rm{mJy\,beam^{-1}}$. SHiZELS-14 is also one of the sources in the catalogue of bright sub-mm sources detected by SCUBA-2 in the COSMOS field \citep{Simpson2019}. The observed-frame $850\,\micron$ flux density measured there is $5.4\pm1.3\,\rm{mJy}$.
\subsection{Astrometric alignment}\label{sec:astrometry}
Accurate astrometric alignment is critical when comparing multi-wavelength emission on small angular scales. However, due to the small fields of view of both the SINFONI ($\sim3''\times3''$) and ALMA ($\sim30''$ diameter) data, aligning the images is non-trivial. Here, we describe the alignment of the images. \\
\indent The ALMA image is expected to be tied to the International Celestial Reference System (ICRS). Although calibration errors and self-calibration processes can lead to astrometric offsets, this is unlikely to be larger than the pixel level ($0.06''$). The JVLA data should also be on the ICRS, and the spatially coincident emission seen by the JVLA and ALMA (Figure \ref{fig:alma_reductions}, right-hand panel), give us confidence in the astrometry of both.\\
\indent We then align all other images to the ICRS. The SINFONI $\rm{H}\alpha$ image was aligned to the same reference frame as the main wide-field HiZELS survey images. We used a broad-band-subtracted narrow-band image from HiZELS, which had been aligned to the Two Micron All-Sky Survey (2MASS), which itself uses the ICRS. We shifted the $\rm{H}\alpha$ image obtained from the SINFONI cube by sub-pixel quantities, and convolved down to the resolution of the broad-band image. Subtracting the images enabled a $\chi^{2}$ fit to define the optimal alignment. Based on these comparisons, we are able to achieve an accuracy on the SINFONI image alignment of $\sim0.2''$. \\
\indent We calibrated the astrometry of the {\it{HST}} images by aligning to  HSC-DR2 \citep{Aihara2019}, which inherits its astrometric accuracy from {\it{Gaia}}. Sources were extracted from the HSC $Y$ and $R$-band images, as well as the two {\it{HST}} images, using the SE{\small{XTRACTOR}} software \citep{Bertin1996}. We matched sources detected in the HSC $Y$-band and the {\it{HST}} F140W (and then the $R$-band and the {\it{HST}} F606W), and constructed histograms of the small offsets between their RA and Dec positions. The peaks of these histograms were selected as the offset to be applied to both the {\it{HST}} images (the same procedure was used in the companion paper, \citealt{Cheng2020}). We also performed this process using catalogued 2MASS sources, and derived essentially identical results. Based on this, and the narrow widths of both histograms, we estimate that the alignment is correct to within $\sim0.1''$. Inspecting our images, gives us confidence in the alignment. As shown in the middle panel of Figure \ref{fig:alma_reductions}, there is faint ALMA flux in the regions that show extended F140W flux. The F140W image also aligns with the SINFONI image in terms of both area covered and areas where the flux peaks.

\begin{figure*}
\centering
\includegraphics[scale=1.0]{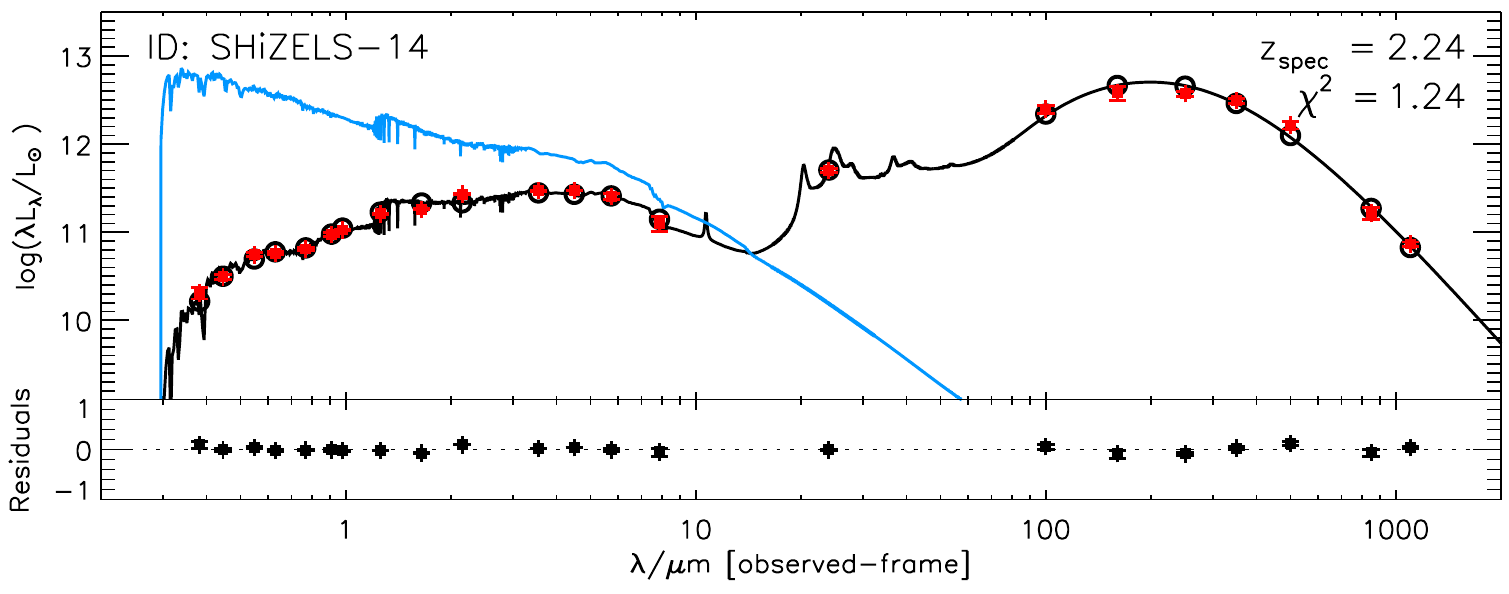}
\caption[SHiZELS-14 SED fit using MAGPHYS]{Data presented in Table \ref{Table:hizels_laigle}, fitted with {\small{MAGPHYS}} \citep{DaCunha2008}. The red points are the observational data, and the open black circles show the model results. The blue line shows the intrinsic stellar SED for the best-fitting model, and the black line shows the SED after dust reprocessing. Residuals are shown in the lower panel. The fitting yields $\rm{SFR}=690\pm30\,M_{\odot}\rm{yr}^{-1}$, $\log_{10}M_{*}/\rm{M_{\odot}}=11.2\pm0.1$, $\log_{10}M_{\rm{dust}}/\rm{M_{\odot}}=8.9\pm0.1$ $A_{v}=1.9\pm0.1$ and $\log_{10}(L_{\rm{TIR}}/\rm{L_{\odot}})= 12.85\pm0.01$.}
\label{fig:magphys_sed}
\end{figure*}

\subsection{Morphologies of astrometrically-calibrated images}\label{sec:morph_comp}
Figure \ref{fig:four_images} shows our four spatially resolved maps after these small astrometric corrections were applied. The emission in all bands is aligned along the same axis. However, the peak of the dust emission probed by ALMA (and confirmed by the $4-8\rm{GHz}$ JVLA imaging) is clearly offset from the peaks of the FUV and $\rm{H}\alpha$ emission. These offsets are far larger than the residual astrometric uncertainties ($0.1-0.2''$). The dust emission is centrally concentrated, whereas there are a number of $\rm{H}\alpha$ peaks along the extended region where dust emission is faint. There is a peak in the emission from both {\it{HST}} bands towards the North-East of the image, yet no detectable dust emission. This is in line with the excess in the CFHT $u^{*}$-band emission (compared to the longer wavelength bands) shown in Figure \ref{fig:cfht_mosaics}. Such offsets are seen in observed dusty galaxies \citep[e.g.][]{Chen2015,Chen2017}, and also in simulations \citep{Cochrane2019}.

\section{Global properties of SHiZELS-14}\label{sec:global}
Before examining the resolved structures of SHiZELS-14 further, we place these into context by deriving the global properties of the galaxy. 
\subsection{SED fitting with {\small{MAGPHYS}}}\label{sec:magphys}
Spectral Energy Distribution (SED) fitting provides a powerful basis for estimating galaxy properties from photometry. 
The {\small{MAGPHYS}} energy balance SED fitting code \citep{DaCunha2008,Cunha2015,Battisti2019} was used to derive physical parameters in \cite{Cheng2020}. We provide details of the fitting here. {\small{MAGPHYS}} employs an energy balance method to match the attenuation of the stellar emission in the UV/optical by dust, and the re-radiation of this energy in the far-infrared. The code uses the stellar population models of \cite{Charlot2003}, with a Chabrier IMF \citep{Chabrier2003} and metallicities between 0.2 and 2 times solar. The star-formation history (SFH) is parametrised as a continuous delayed exponential function and to reproduce starbursts, {\small{MAGPHYS}} also adds bursts to the star-formation history.  Dust attenuation is modelled using two components following \cite{Charlot2000}. The code was extensively tested with both observational constraints on SMGs and against model star-forming galaxies from the EAGLE simulation by \cite{Dudzeviciute2019} and shown to perform well for these highly dust-obscured galaxies.\\
\indent We fit the photometry presented in Table \ref{Table:hizels_laigle}.
We estimate $\log_{10}M_{*}/\rm{M_{\odot}}=11.2\pm0.1$, $\log_{10}M_{\rm{dust}}/\rm{M_{\odot}}=8.9\pm0.1$, and $\rm{SFR}=690\pm30\,\rm{M_{\odot}yr^{-1}}$ (see Table \ref{Table:hizels_fp}). The estimated total infrared luminosity is  $\log_{10}(L_{\rm{TIR}}/\rm{L_{\odot}})= 12.85\pm0.01$, and the estimated dust attenuation in the $V$-band is $A_{v}=1.9\pm0.1$. We obtain consistent results using another SED fitting code, {\small{BAGPIPES}} (see appendix). We have also used {\small{BAGPIPES}} to experiment with different SFH parametrisations, which yield very similar fits to the photometry and consistent values for stellar mass. All parametrisations, even those allowing multiple bursts, favour a recent (at $z=2.24$), rapid burst of star formation in which the vast majority of the stellar mass is formed.

\begin{table}
\begin{center}
\begin{tabular}{l|c|c}
Basic property & Measurement & Reference \\ 
\hline
RA (J2000) & 10:00:51.6 & Swinbank+12\\
Dec (J2000) & +02:33:34.5 & Swinbank+12\\
$z_{\rm{H}\alpha}$ & 2.2418 & Swinbank+12\\
\hline
Derived property & Measurement & Reference \\ 
\hline
$\log_{10}M_{*, \rm{SED}}/\rm{M_{\odot}}$ & $11.2\pm0.1$ & This paper \\
$\log_{10}M_{\rm{gas}}/\rm{M_{\odot}}$ & $10.1\pm0.4$ & Swinbank+12\\
$\log_{10}M_{\rm{dust}}/\rm{M_{\odot}}$ & $8.9\pm0.1$ & This paper \\
$\log_{10}L_{\rm{TIR}}/\rm{L_{\odot}}$ & $12.85\pm0.01$ & This paper\\
$\rm{SFR}_{SED}/\rm{M_{\odot}yr^{-1}}$ & $690\pm30
$ & This paper \\
$R_{e,\,\rm{H}\alpha}/\rm{kpc}$ & $4.6\pm0.4$ & Swinbank+12\\
$R_{e,\,\rm{opt}}^{\rm{maj}}/\rm{kpc}$ & $4.6\pm0.2$ & This paper \\
$R_{e,\,\rm{FIR}}^{\rm{maj}}/\rm{kpc}$ & $4.5\pm0.2$ & This paper \\
$A_{v}$ & $1.9\pm0.1$ & This paper \\
\hline
\end{tabular}
\caption{Summary of properties of SHiZELS-14. Full details of SFRs derived using different methods are presented separately in Table \ref{table:sfrs}.}
\label{Table:hizels_fp}
\end{center}
\end{table}

\subsection{Fitting the dust SED}\label{sec:dust_sed}
\begin{figure}
\includegraphics[scale=0.56]{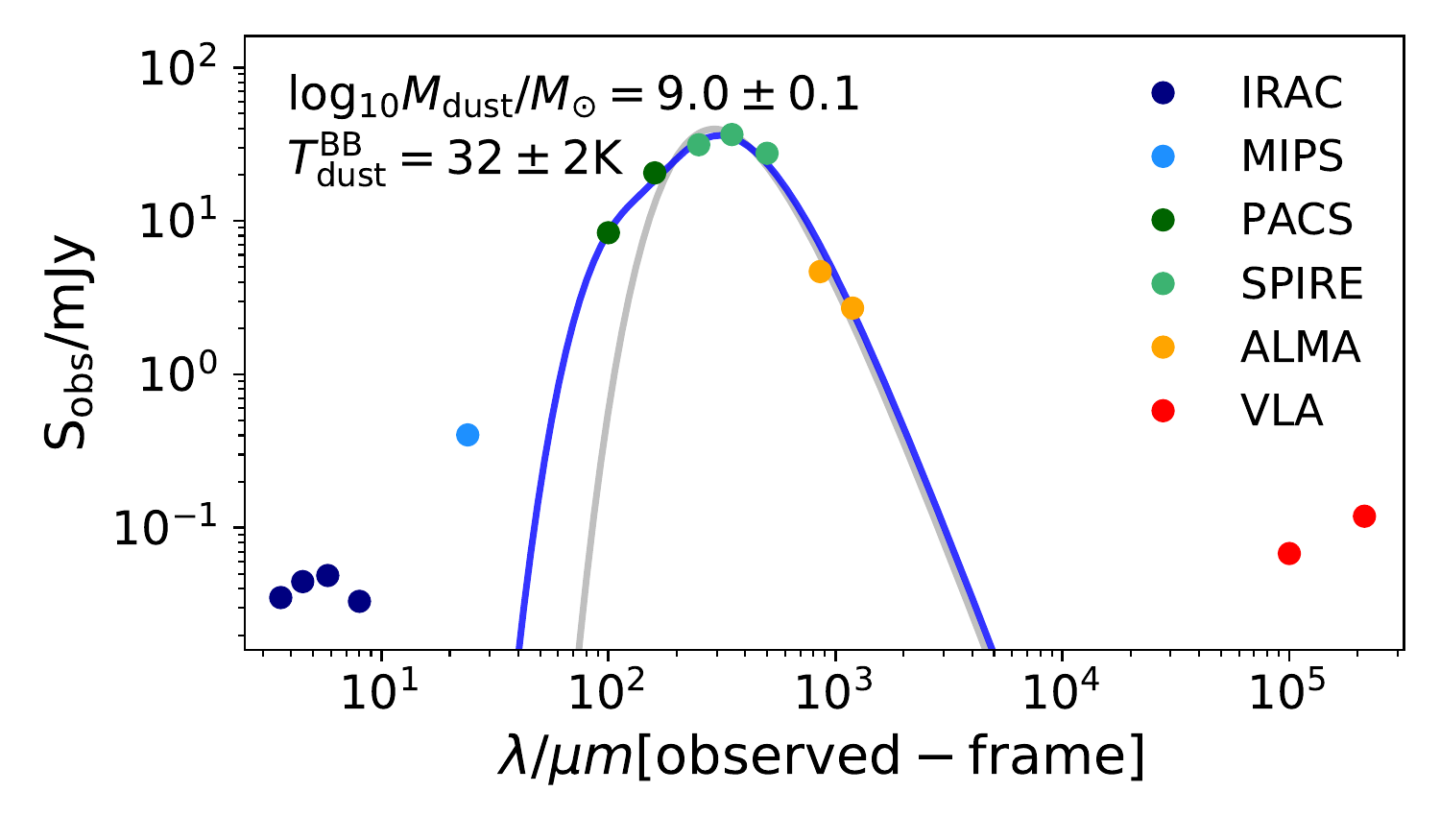}

\caption[Fit to SHiZELS-14's dust SED]{The dust SED of SHiZELS-14, constructed using collated archival data and the new ALMA $260\rm{GHz}$ data. Error bars are plotted on the data points, but are small. A two grey body model parametrisation (blue fitted curve) provides a good fit to both the cold and warm dust components. Integrating the $8-1000\,\micron$ emission gives $\log_{10}(L_{\rm{TIR}}/\rm{L_{\odot}})= 12.81\pm0.02$ and SFR$_{8-1000\,\mu\rm{m}} = 950\pm50\,\rm{M_{\odot}}\rm{yr}^{-1}$. A single component model (grey fitted curve) is unable to fit the $100\,\mu\rm{m}$ PACS data. The characteristic dust temperature derived from this fit ($T_{\rm{dust}}^{\rm{BB}}=32\pm2\,\rm{K}$) enables us to constrain the dust mass ($\log_{10}M_{\rm{dust}}/\rm{M_{\odot}} = 9.0\pm0.1$).}
\label{fig:dust_sed_fit}
\end{figure}

To assess the sensitivity of the derived {\small{MAGPHYS}} parameters in the far-infrared, we also fit the MIR-to-FIR SED of SHiZELS-14 separately, using only data from ALMA and {\it{Herschel}}. We parametrise the emission from cold and warm dust using a simple two-body model: \\
\begin{equation}
f_{\nu}(\rm{mJy}) = A_{\rm{warm}}\lambda^{-\beta_{warm}}B_{\nu}(T_{\rm{warm}}) + A_{\rm{cold}}\lambda^{-\beta_{cold}}B_{\nu}(T_{\rm{cold}}) ,
\end{equation}\\
where $A_{\rm{warm}}$ and $A_{\rm{cold}}$ are normalisations and $B_{\nu}(T)$ is the Planck function, from dust grains radiating at rest frequency $\nu$, at temperature T. All wavelengths were input at their rest-frame. In line with the literature, we have fixed $\beta=2$ for both the cold and warm dust components, to minimise the number of fitting parameters. We use the {\small{EMCEE}} MCMC python package \citep{Foreman-Mackey2016}, with 300 walkers and 5000 steps. This yields posterior estimates: $\log_{10}A_{\rm{warm}} = 5.4\pm0.3$, $T_{\rm{warm}} = 64\pm6\,\rm{K}$, $\log_{10}A_{\rm{cold}} = 7.6\pm0.1$, and $T_{\rm{cold}} = 28\pm2\,\rm{K}$. The best-fitting model is shown in Figure \ref{fig:dust_sed_fit}. Note that there is a known strong degeneracy between $\beta_{\rm{cold}}$ and $T_{\rm{cold}}$, and a 5-parameter fit that allows $\beta_{\rm{cold}}$ to vary favours a higher $\beta_{\rm{cold}}$ and a lower $T_{\rm{cold}}$.\\
\indent To derive the dust temperature in a consistent way to other studies in the literature, we additionally fit a single modified black body model to the data, also shown in Figure \ref{fig:dust_sed_fit}. We derive a characteristic dust temperature of $T_{\rm{dust}}^{\rm{BB}}=32\pm2\,\rm{K}$, in good agreement with the values derived by \cite{Dudzeviciute2019} for their sample of SMGs. However, this single component model struggles to fit the $100\,\mu\rm{m}$ PACS data point, and it is necessary to boost the errors on that point artificially to get a good fit to the longer wavelength data. This suggests a contribution from hotter dust, perhaps indicative of an obscured AGN. We will discuss this further in Section \ref{sec:agn}.

\subsection{Calculation of cold dust mass and TIR luminosity}\label{sec:mcmc_dust_mass}
Assuming the dust is optically thin at the rest-frame frequency, the dust mass is given by (e.g. \citealt{James2002}): 
\begin{equation}
M_{\rm{dust}} = \frac{1}{1+z} \frac{S_{\rm{obs}}D_{L}^{2}}{\kappa_{\rm{rest}}B_{\nu}(T_{\rm{dust}}^{\rm{BB}})},
\end{equation}
where $S_{\rm{obs}} = 2.7\,\rm{mJy}$ is the observed flux density of the source at $260\,\rm{GHz}$, $D_{L}$ is the luminosity distance, $\nu=836\,\rm{GHz}$ is the rest-frame frequency, $\kappa_{\rm{rest}}$ is the mass absorption coefficient at this frequency, and $T_{\rm{dust}}^{\rm{BB}}=32\pm2\,\rm{K}$ is the characteristic temperature derived from the single component modified black body fit. We used $\kappa_{850} = 0.07\pm0.02\,\rm{m}^{2}\rm{kg}^{-1}$ \citep{James2002}, which gives $\log_{10} M_{\rm{dust}}/\rm{M_{\odot}} = 9.0\pm0.1$ (in good agreement with the value derived from {\small{MAGPHYS}} fits, $\log_{10} M_{\rm{dust}}/\rm{M_{\odot}} = 8.9\pm0.1$). These estimates are consistent with a high dust-to-stellar mass ratio, $\log_{10}(M_{\rm{dust}}/M_{*}) = -2.2\pm0.2$, which is comparable to the ratios derived by \cite{Calura2017} for SMGs of stellar mass $\sim10^{11}\,\rm{M_{\odot}}$ at $z\sim1-3$ (see also Figure \ref{fig:smg_plots}). We also integrate the two-body fits at wavelengths $8-1000\,\micron$ within the MCMC fit (enabling us to fold in the correlations between fitted parameters), obtaining an estimate for the total IR luminosity, $\log_{10}(L_{\rm{TIR}}/\rm{erg\,s^{-1}})=46.39\pm0.02$, and $\log_{10}(L_{\rm{TIR}}/\rm{L_{\odot}})= 12.81\pm0.02$. The TIR-based SFR is $950\pm50\,\rm{M_{\odot}}\rm{yr}^{-1}$ (using the \citealt{Kennicutt2012} SFR calibration, with a \citealt{Kroupa2002} IMF).\\

\begin{table*}
\begin{center}
\begin{tabular}{l|l|c}
Waveband (Instrument) & Formula for $\log_{10}(\rm{SFR}/M_{\odot}\rm{yr}^{-1})$ & $\rm{SFR}/M_{\odot}\rm{yr}^{-1}$ \\ 
\hline
SFRs from individual tracers & & \\
\hline
${\rm\bf{TIR_{8-1000\micron}}}$ \bf{(dust SED fit)} & $\log_{10}(L_{\rm{TIR}}/\rm{erg\,s^{-1}}) - 43.41$ & $\mathbf{950 \pm 50}$ \vspace{0.1cm}\\
Radio (1.4\,GHz, VLA, \citealt{Bell2003a} conversion) & $\log_{10}(L_{1.4\rm{GHz,\,rest}}/\rm{erg\,s^{-1}\,Hz^{-1}}) - 28.43$ &  $1180\pm100$\\
Radio (1.4\,GHz, VLA, \citealt{Kennicutt2012}) & $\log_{10}(L_{1.4\rm{GHz,\,rest}}/\rm{erg\,s^{-1}\,Hz^{-1}}) - 28.2$ &  $2010\pm170$\\ 
$\rm{H}\alpha$ (SINFONI/VLT) & $\log_{10}(L_{\rm{H}\alpha}/\rm{erg\,s^{-1}}) - 41.27$ & $33\pm2$ \\
'' & $L_{\rm{H}\alpha}$ corrected using $1\,$mag dust extinction & $83\pm5$ \\
'' & $L_{\rm{H}\alpha}$ corrected using $M_{*}$-dependent dust extinction & $180\pm10$ \\ 
& of \cite{Garn2010a}, $\log_{10}M_{*}/\rm{M_{\odot}} = 11.2$ & \\
'' & $L_{\rm{H}\alpha}$ corrected using $A_{\rm{H}\alpha}=1.6\pm0.1$, derived from scaled $A_{V}$ & $140\pm20$ \\
'' & $L_{\rm{H}\alpha}$ corrected using $A_{\rm{H}\alpha}=3.7\pm0.1$, derived from scaled $A_{V}$ \& & $1000\pm100$\\
'' &  preferential extinction of birth clouds &  \\

FUV ({\it{HST}} F606W) & $\log_{10}(\nu L_{\nu}/\rm{erg}\,\rm{s}^{-1}) - 43.17$ & $13\pm1$ \\
'' & corrected using $A_{1600}$ derived from $\beta$, with $\beta=-0.5\pm0.1$ & $300^{+70}_{-50}$ \\
'' & corrected using $A_{\rm{UV}}=4.3\pm0.2$, derived from scaled $A_{V}$ & $680\pm130$ \\
\hline
SFRs from combinations of tracers & & \\
\hline
FUV + TIR & $L_{\rm{\nu,corr}} = L_{\rm{\nu,obs}} + 0.27L_{\rm{TIR}}$, $L_{\nu}-\rm{SFR}$ conversion above & $440\pm20$ \\ 
$\rm{H}\alpha$ + TIR & $L_{\rm{H}\alpha,corr} = L_{\rm{H}\alpha,obs} + 0.0024L_{\rm{TIR}}$, $L_{\rm{H}\alpha}-\rm{SFR}$ conversion above & $330\pm20$ \\ 
FUV + radio & $L_{\rm{FUV,corr}} = L_{\rm{FUV,obs}} + 4.2\times 10^{14}L_{1.4\rm{GHz}}$ & $990\pm80$ \\
\hline
SFRs from SED fitting & & \\
\hline
{\small{MAGPHYS}} & & $690\pm30$ \\
{\small{BAGPIPES}} & & $660\pm60$ \\
\hline

\end{tabular}
\caption[The global SFR of SHiZELS-14, derived from different combinations of SFR indicators]{The global SFR of SHiZELS-14, derived from different combinations of SFR indicators, using a \cite{Kroupa2002} IMF. Inferring an SFR from dust-uncorrected fluxes at short-wavelengths yields SFRs of $<50 \rm{M_{\odot}\rm{yr}^{-1}}$. These UV or $\rm{H}\alpha$-inferred SFRs lie well below the values derived from the TIR or radio ($\rm{SFR}= 1000-2000\,\rm{M_{\odot}yr^{-1}}$). Applying either a standard dust correction corresponding to $A_{\rm{H}\alpha}=1$ or a stellar mass dependent dust correction provides only a modest increase in $\rm{H}\alpha$-derived SFR ($\rm{SFR}=100-200\,\rm{M_{\odot}\rm{yr}^{-1}}$). Similarly, correcting the UV-derived SFR using the IRX-$\beta$ relation derived using HST data only raises the SFR to $300\, \rm{M_{\odot}\rm{yr}^{-1}}$. We also estimate $A_{\rm{H}\alpha}$ and $A_{1600}$ by scaling the $A_{V}$ derived from {\small{MAGPHYS}} according to a \cite{Calzetti2000} law. This correction brings the UV-derived SFR into better agreement with the  {\small{MAGPHYS}}-derived SFR, however the $\rm{H}\alpha$-derived SFR remains low, indicating additional extinction of the $A_{\rm{H}\alpha}$ line. Including additional extinction of the $A_{\rm{H}\alpha}$ line according to \cite{Charlot2000} brings the $\rm{H}\alpha$-derived SFR into line with the FIR estimate ($\rm{SFR}\sim1000\,\rm{M_{\odot}yr^{-1}}$).}
\label{table:sfrs}
\end{center}
\end{table*}

\subsection{The inappropriateness of the $\rm{IRX}-\beta$ relation}\label{sec:global_irx-beta}
The $\rm{IRX}-\beta$ relation \citep{Calzetti1994,Meurer1999} between the ratio of the FIR and UV luminosity ($\rm{IRX} = L_{\rm{FIR}}/L_{1600}$) and the spectral slope ($\beta$, where $f_{\lambda}\propto \lambda^{\beta}$) evaluated at $1600\,\angstrom$ is a popular method used to infer SFRs where only  rest-frame UV luminosities are available. This appears to work for samples of galaxies with relatively low dust content (especially at very high redshift). However, individual galaxies show a large amount of scatter around this relation, and it has been shown that this method is not appropriate highly star-forming galaxies \citep[e.g.][]{Casey2014,Chen2017,Narayanan2018}, although it is difficult to identify these based on their UV properties.\\
\indent We can derive both IRX and $\beta$ for SHiZELS-14. We use the publicly available {\it{HST}} $I_{814}$-band image ($\lambda_{\rm{mean}}=8100\,\angstrom$, rest-frame $\lambda_{\rm{mean}}=2500\,\angstrom$), along with our own F606W images ($\lambda_{\rm{mean}}=6000\,\angstrom$, rest-frame $\lambda_{\rm{mean}}=1850\,\angstrom$), to calculate $\beta$. Adopting our derived $\beta=-0.5\pm0.1$, and applying the relation $A_{1600}=4.43+1.99\beta$, we derive $A_{1600}=3.4\pm0.2$ (lower than the $A_{1600}$ estimated from scaling the $A_{V}$ obtained via SED fitting; see Table \ref{table:sfrs}). Correcting the global SFR inferred from the FUV flux accordingly yields $\rm{SFR}=300^{+70}_{-50}\,M_{\odot}\rm{yr^{-1}}$. This estimate is approximately two times lower than the SFR inferred from the SED fitting presented in Sections \ref{sec:magphys} and \ref{sec:dust_sed}, and three times lower than the TIR-derived SFR in Section \ref{sec:mcmc_dust_mass}. We calculate IRX using the TIR luminosity derived in Section \ref{sec:mcmc_dust_mass}, and the rest-frame $1851\,\angstrom$ luminosity. Globally, the galaxy has $\log_{10}\rm{IRX}=2.09\pm0.06$. In combination with the derived $\beta$, this places it $\sim0.7\pm0.1\,\rm{dex}$ above the \cite{Meurer1999} relation. This highlights that the galaxy has a higher TIR luminosity than expected from the derived UV slope, or, equivalently, a much shallower UV slope than would be expected given the radio of the global TIR luminosity to UV luminosity. This is likely to be because the UV and FIR emission are not co-located, as shown in Figure \ref{fig:four_images}. Because of this, the UV slope is measured from the UV emission escaping from one region of the galaxy, which is not where most of the FIR emission arises \citep[see also][]{Goldader2002,Cochrane2019}. SHiZELS-14 highlights that the IRX relation does not provide reliable estimates of the FIR emission for the most dusty galaxies, as also argued by \cite{Chen2017}. 

\subsection{Global SFR estimation}\label{sec:global_sfrs}
In Table \ref{table:sfrs} we present global SFR estimates from global measurements in different wavebands, using the calibrations of \cite{Kennicutt2012} and assuming a \cite{Kroupa2002} IMF. It is clear that applying standard SFR calibrations to flux measurements that probe star formation via direct emission at shorter wavelengths predict vastly lower SFRs than the dust-obscured tracers. This suggests that the deficit in global SFR derived from the dust-sensitive SFR indicators is due to the highly dusty nature of this galaxy. In the following section, we explore the differences in the spatially-resolved SFRs, derived at different wavelengths.

\subsection{The lack of evidence for AGN activity}\label{sec:agn}
As discussed in Section \ref{sec:global_sfrs}, the SFRs derived from the UV, $\rm{H}\alpha$ and FIR differ greatly. In this section, we investigate whether the presence of an active galactic nucleus (AGN) could be a factor in this. In this scenario, the extreme dust continuum emission towards the centre of the galaxy could be powered by heating from a central AGN, rather than a compact region of star formation. Since different types of AGN emit in different wavebands (see \citealt{Heckman2014} for a review, and \citealt{Garn2010a} for a discussion of AGN within the HiZELS sample), identification of AGN requires a multi-wavelength approach. Here, we use some of the key methods for AGN identification to hunt for signs of AGN activity.

\subsubsection{No X-ray detection}
X-ray emission probes the accretion disk corona very close to a supermassive black hole. The {\it{Chandra}} COSMOS-Legacy Survey \citep{Civano2016} imaged $2.2\,\rm{deg}^{2}$ of the COSMOS field in the wavelength range $0.5-10\,\rm{keV}$. SHiZELS-14 lies in the central region of the COSMOS-Legacy field, where effective exposure times are $\sim160\,\rm{ks}$. The limiting depths are $2.2\times10^{-16}$, $1.5\times10^{-15}$, and $8.9\times10^{-16}$ $\rm{erg\,cm^{-2}\,s^{-1}}$ in the bands $0.5-2$, $2-10$, and $0.5-10\, \rm{keV}$, respectively. At these limiting depths, SHiZELS-14 is undetected. We derive a limit on the rest-frame hard-band $2-10\,\rm{keV}$ luminosity following \cite{Alexander2003}: $L_{2-10,\rm{lim}} =  4\pi D_{L}^{2} f_{2-10,\rm{lim}} (1+z)^{\Gamma-2}$, using $\Gamma=1.9$ \citep{Nandra1994}. This gives $L_{2-10}<5.1\times10^{43}\,\rm{erg\,s^{-1}}$. \\
\indent We can predict the X-ray luminosity associated with star formation using the $L_{2-10}$-SFR calibration proposed by \cite{Kennicutt2012} and the SFR measured from the other indicators. Given the SFR derived from the dust SED fit, $950\pm50\,\rm{M_{\odot}yr^{-1}}$, we estimate $L_{2-10}=(5.6\pm0.3)\times10^{42}\,\rm{erg\,s^{-1}}$. This is an order of magnitude lower than the limit imposed by the survey depth. Therefore, the lack of an X-ray detection is consistent with SHiZELS-14 being a star-forming galaxy.

\subsubsection{No [N{\small{II}}]/$\rm{H}\alpha$ excess}
The ratio of [N{\small{II}}]-to-$\rm{H}\alpha$ line flux reflects the hardness of the ionising source driving the nebular emission, and hence can be used to infer the presence of an AGN, often in combination with other line ratios \citep[e.g.][]{Baldwin1981}. For SHiZELS-14, there is no evidence for a strong excess in [N{\small{II}}]/$\rm{H}\alpha$. For the clump nearest the peak of the rest-frame FIR emission, [N{\small{II}}]/$\rm{H}\alpha=0.12$ \citep[][clump 14a]{Swinbank2012b}, well within the range expected for star-forming regions \citep[e.g.][]{Kewley2006}. \cite{Swinbank2012a} show that the [N{\small{II}}]/$\rm{H}\alpha$ radial profile of SHiZELS-14 is slightly negative, in line with the rest of the SHiZELS sample. The derived gradients reflect slightly enhanced metallicity towards the central regions of the SHiZELS galaxies, consistent with simulations of star-forming galaxies of similar mass and redshift \citep[e.g.][]{Ma2017a}.  

\begin{figure*}
\includegraphics[scale=0.5]{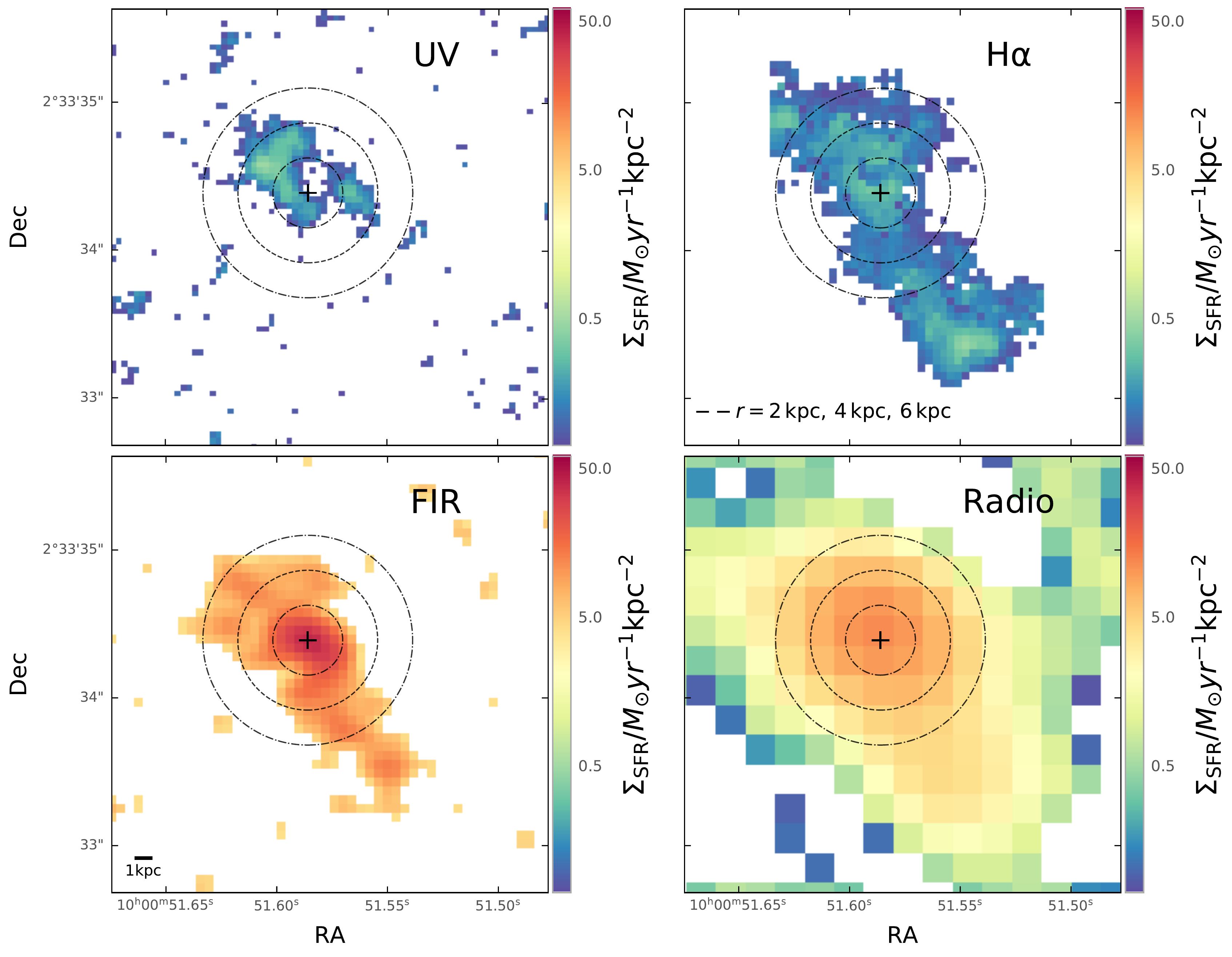}
\caption[Maps of SFR surface density, derived for each of the four SFR indicators]{Maps of SFR surface density, derived for each of the four SFR indicators using the luminosity-SFR calibrations of \cite{Kennicutt2012} (upper two and lower left panels) and \cite{Bell2003a} (lower right panel). Note that dust corrections were not applied to the UV or $\rm{H}\alpha$ maps. Pixels that have fluxes below the minimum of our $\Sigma_{\rm{SFR}}$ scale ($0.07\,\rm{M_{\odot}}\rm{yr}^{-1}\rm{kpc}^{-2}$) are coloured white to avoid overly noisy images. We plot the maps on the same SFR scale, to compare the SFRs directly, and show the position of the peak of the ALMA emission as a black cross on each panel. We also overplot three concentric rings, of radii $2\,\rm{kpc}$, $4\,\rm{kpc}$ and $6\,\rm{kpc}$. It is clear that the derived SFRs differ across the spatial extent of the galaxy, not only in its dusty centre. The UV map shows a 'hole' where the rest-frame FIR emission peaks. As shown in Figure \ref{fig:alma_reductions}, the angular resolution of the radio imaging is lower than the other images, which causes the emission to appear more extended.}
\label{fig:sfr_maps}
\end{figure*}

\subsubsection{No mid-infrared excess}
Obscured AGN are characterised by a strong mid-infrared (rest-frame $\sim3-30\,\mu\rm{m}$) excess, produced by a dusty obscuring torus. Our {\small{MAGPHYS}} fit (Figure \ref{fig:magphys_sed}) shows no sign of such an excess, being well-fitted by an SED constructed without AGN templates. Fitting the SED with {\small{CIGALE}}, which does allow for the inclusion of emission from AGN, provides no evidence of an AGN ($f_{\rm{AGN,\,best}}=0.001$). In addition to this, the characteristic temperature derived from fits to the dust SED (which is well-constrained due to the known redshift) is $32\pm2\,\rm{K}$, well within the normal range for star-forming galaxies. However, a single modified black body fails to fit the shortest wavelength PACS data point, indicating non-negligible emission from hotter dust. This could hint at some contribution to the FIR emission from an obscured AGN, though there are numerous examples of hot dust associated with star formation \citep[e.g.][]{Magnelli2014,Faisst2020}.

\subsubsection{Position on the IR-radio relation consistent with star formation}\label{sec:ir-radio}
The ratio of IR to radio luminosity \citep[e.g.][]{Appleton2004} is frequently employed to separate radio-loud AGN from star-forming galaxies. Following \cite{Ivison2010}, we use the following equation with the TIR luminosity calculated in Section \ref{sec:mcmc_dust_mass}:

\begin{equation}
    q_{\rm{TIR}}=\log\Bigg(\frac{L_{\rm{TIR}}}{3.75\times10^{12}\rm{W}}\Bigg)-\log\Bigg(\frac{L_{1.4\rm{GHz\,rest}}}{\rm{WHz^{-1}}}\Bigg).
\end{equation}
The rest-frame $1.4\rm{GHz}$ luminosity is:

\begin{equation}
\begin{multlined}
L_{1.4\rm{GHz,\,rest}} = \frac{4\pi D_{L}^{2}}{(1+z)^{1+\alpha}} \Bigg(\frac{\nu_{1.4\rm{GHz}}}{\nu_{\rm{obs}}}\Bigg)^{\alpha}S_{1.4\rm{GHz,\,obs}}\\
=10^{24.54\pm0.08}\,\rm{WHz^{-1}}.
\end{multlined}
\end{equation}\\
We assume a spectral index $\alpha=-0.77$, derived from the VLA $3\,\rm{GHz}$ and $1.4\,\rm{GHz}$ data. This gives $q_{\rm{TIR}}=2.28\pm0.10$. This is broadly in line with the distribution of $q_{\rm{TIR}}$ values for the $250\,\micron$-selected sample of \cite{Ivison2010} (median $q_{\rm{TIR}} = 2.4$, $\sigma_{q} = 0.24$; see also \citealt{Algera2020}).    
The $q_{\rm{TIR}}$ value for SHiZELS-14 is well within $1\sigma$ of the median relation derived for star-forming galaxies. This indicates that the radio continuum emission is not contaminated by a compact radio core. Overall, we find no evidence that SHiZELS-14 is host to a radio-loud AGN.

\section{Resolved star-formation rates and dust attenuation}\label{sec:resolved_comparison}

\begin{figure}[!t]
\centering
\includegraphics[scale=0.9]{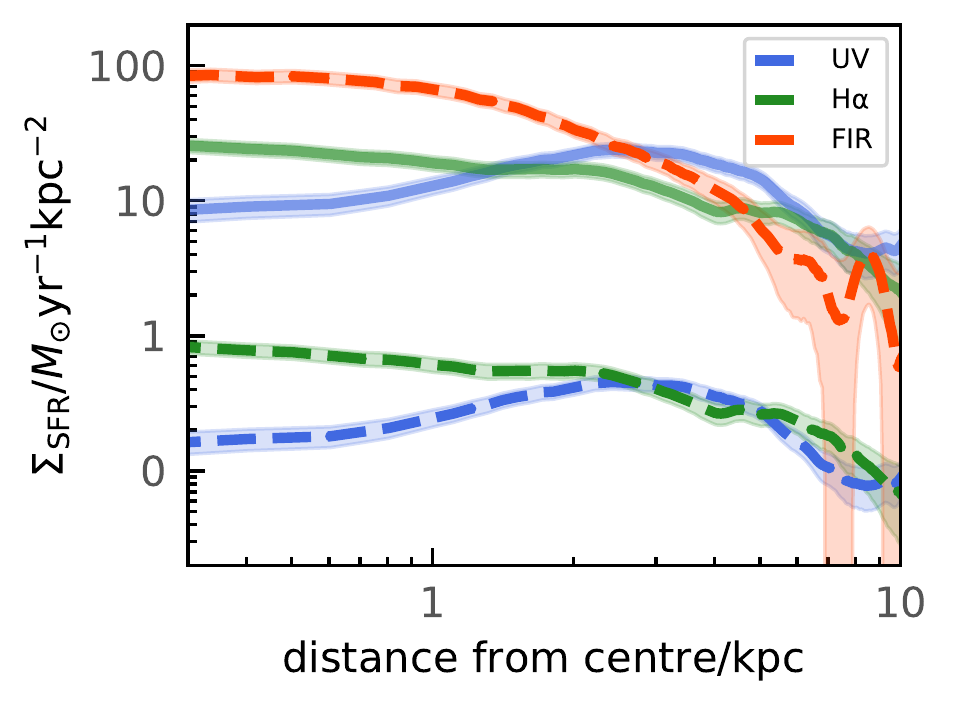}
\caption[Star-formation rate surface density profiles]{Star-formation rate surface density profiles derived using rest-frame FUV F606W, $\rm{H}\alpha$, and rest-frame FIR flux map (scaled to the SFR derived from fits to the dust SED; note that this assumes a constant dust temperature across the galaxy). The profiles are centred on the peak of the rest-frame FIR emission, shown by a black cross in Figure \ref{fig:sfr_maps}. The thick dashed lines show the surface densities derived using \cite{Kennicutt2012} calibrations, with no dust corrections applied. The solid transparent lines show the profiles derived using an $A_{\rm{H}\alpha}=3.7$ and $A_{\rm{UV}}=4.3$, derived using $A_{V}=1.9$ from {\small{MAGPHYS}}, the attenuation curve of \cite{Calzetti2000} and the preferential attenuation towards birth clouds proposed by \cite{Charlot2000}. These corrections can bring the profiles broadly into line at large radii, but still underestimate the star-formation rate surface density at radii less than $\sim2\,\rm{kpc}$, where the rest-frame FIR emission peaks.}
\label{fig:radial_profiles}
\end{figure}

\subsection{Resolved star-formation rates}\label{sec:resolved_sfrs}
In Figure \ref{fig:sfr_maps}, we present maps of SFR surface density, derived for each of the four SFR tracers using the luminosity-SFR calibrations of \cite{Kennicutt2012} and \cite{Bell2003a}. In order to do this, we assume that these global calibrations are also valid on smaller spatial scales, which may not be the case. In reality, gradients in dust temperature and opacity \citep[e.g.][]{Galametz2012} may apply to the TIR model, and gradients in the reddening will influence the $\rm{H}\alpha$ and UV maps. The radio emission is sensitive to cosmic ray propagation and starburst age, which likely results in smoother and more extended emission than the true SFR distribution \citep{Thomson2019}. Making this assumption and adopting standard SFR calibrations, it is clear from Figure \ref{fig:sfr_maps} that the SFRs derived from the four indicators differ across the galaxy. To investigate this more quantitatively, we derive star-formation rate radial profiles by applying \cite{Kennicutt2012} calibrations to the rest-frame FUV F606W, $\rm{H}\alpha$, and TIR flux maps (see Figure \ref{fig:radial_profiles}, thick dashed lines). The three profiles are discrepant, with the TIR-based SFR profile increasing sharply towards the centre, and the FUV-derived profile decreasing at radii smaller than $\sim2\,\rm{kpc}$, in line with the 'hole' observed at the position of the peak of the dust continuum emission (see Figure \ref{fig:sfr_maps}). Without any corrections for dust attenuation, the FUV and $\rm{H}\alpha$-derived SFRs are lower than the FIR-derived SFR across the radial extent of the galaxy. The FUV profile broadly follows the $\rm{H}\alpha$ profile in shape, but with a different normalisation. The FUV is most strongly attenuated by dust, and yields the lowest dust-uncorrected SFRs across the galaxy. Thus, the discrepancy between the SFRs derived globally cannot be attributed solely to the compact dusty centre of the galaxy, though this is where the measurements are most discrepant. Instead, short-wavelength light is attenuated across the galaxy. \\
\indent We also show the effects of applying a dust correction. $A_{\rm{H}\alpha}$ and $A_{\rm{UV}}$ are calculated from the {\small{MAGPHYS}}-derived $A_{V}$, according to the \cite{Calzetti2000} law and a \cite{Charlot2000} birth cloud attenuation. These dust corrections bring the outermost regions of the FUV and $\rm{H}\alpha$ profiles further towards agreement at radii greater than $\sim2\,\rm{kpc}$ (see transparent solid lines). However, it is clear that the UV and $\rm{H}\alpha$-derived SFR estimates are much lower than the TIR-derived estimate in the centre, particularly at radii less than $\sim2\,\rm{kpc}$. This reflects strong central star formation and a steep gradient in dust attenuation across the galaxy, which may be even stronger if there is a significant gradient in either dust temperature or opacity.
\begin{figure*}
\includegraphics[width=\columnwidth]{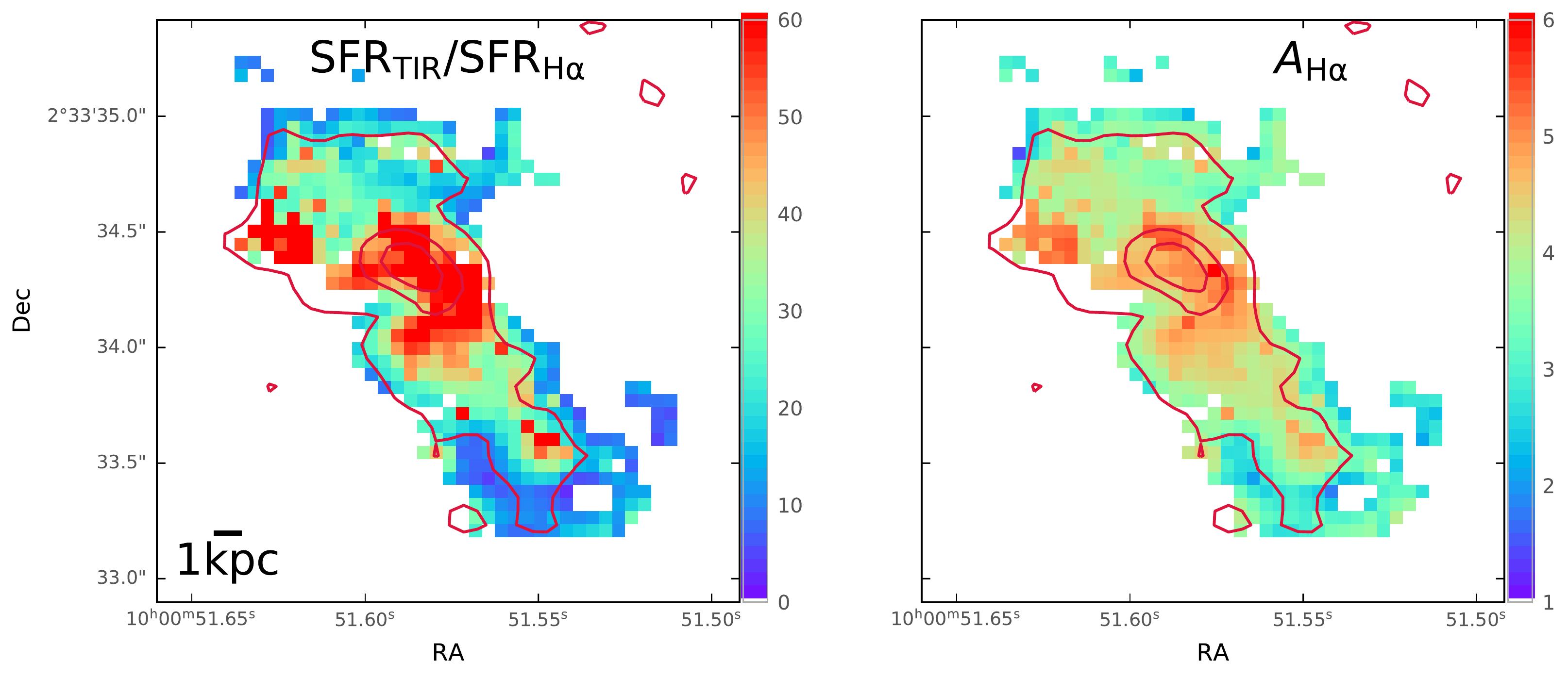}
\caption[Deriving $A_{\rm{H}\alpha}$ from $\rm{H}\alpha$ and TIR maps]{Left: the ratio of TIR-derived SFR to $\rm{H}\alpha$-derived SFR, assuming the luminosity-SFR calibrations of \cite{Kennicutt2012}, without any correction for dust attenuation. The TIR-derived SFR is larger than that derived from $\rm{H}\alpha$ across the full extent of the galaxy, but the estimates are discrepant by a factor of $\sim50$ in the dusty central region. Right: the dust attenuation $A_{\rm{H}\alpha}$ derived from this ratio. Where the $\rm{H}\alpha$ flux is below the detection limit, neither ratio nor $A_{\rm{H}\alpha}$ value is plotted. $A_{\rm{H}\alpha}$ varies across the galaxy, within a broad range $A_{\rm{H}\alpha}\sim2-6$. Surveys such as HiZELS often assume a modest global dust correction of $A_{\rm{H}\alpha}=1$, but the dust attenuation of SHiZELS-14 derived here is well in excess of this value. ALMA contours are overlaid on both panels in red.}
\label{fig:aha_plot}
\end{figure*}

\subsection{Inferring dust attenuation using $\rm{H}\alpha$ and FIR maps}\label{sec:resolved_sfrs}
In Figure \ref{fig:sfr_maps}, we showed that the SFR surface densities derived in different wavebands from dust-uncorrected fluxes of the dust-sensitive tracers are far lower than the TIR measurement. We can use this to estimate the spatially-resolved dust attenuation. In the left-hand panel of Figure \ref{fig:aha_plot}, we present the ratio of the $\rm{H}\alpha$-derived SFR (with no dust correction applied) to the TIR-derived SFR. We can also use this ratio of the fluxes to estimate $A_{\rm{H}\alpha}$ in a spatially-resolved way, as follows. Folding in a dust-correction to the $\rm{H}\alpha$ flux, and then equating the two SFRs:

\begin{equation}
\rm{SFR}/M_{\odot}\rm{yr}^{-1} = L_{\rm{TIR}} \times 10^{-43.41} = L_{\rm{H}\alpha}\times 10^{-41.27}\times 10^{0.4 A_{\rm{H}\alpha}}
\end{equation}
yields an expression for $A_{\rm{H}\alpha}$:
\begin{equation}
A_{\rm{H}\alpha} = 2.5 \log_{10}\Bigg(\frac{\rm{L_{TIR}}}{\rm{L}_{\rm{H}\alpha}}\Bigg)-5.35.
\end{equation}
Note that this method assumes that $\rm{H}\alpha$ and FIR flux are tracing only recently formed stars, and sensitive to star formation on the same timescales. \\
\indent We plot the spatially-resolved $A_{\rm{H}\alpha}$ in the right-hand panel of Figure \ref{fig:aha_plot}. $A_{\rm{H}\alpha}$ substantially exceeds $1$, the canonical value applied to global studies, across the spatial extent of the galaxy. The derived $A_{\rm{H}\alpha}$ is larger than that derived from scaling $A_{V}$ according to the \cite{Calzetti2000} law and \cite{Charlot2000} prescription ($A_{\rm{H}\alpha}=3.7$) in the dustiest parts of the galaxy. In the most dusty central region, it reaches a peak of $A_{\rm{H}\alpha}\sim5$. In fact, the true value is likely to be above that due to gradients in the dust temperature and opacity.

\subsection{Origin of the observed rest-frame UV flux}\label{sec:predict_uv}
While the $\rm{H}\alpha$ emission traces broadly the same spatial extent as the rest-frame FIR emission, the rest-frame UV emission is concentrated towards the North East of the galaxy. Assuming that $\rm{H}\alpha$ and UV are probing the same star formation, we can predict the observed UV flux, $I_{\rm{obs,UV}}$, from the observed $\rm{H}\alpha$ flux, $I_{\rm{obs,H}\alpha}$, using the $A_{\rm{H}\alpha}$ map shown in Figure \ref{fig:aha_plot} and:
\begin{equation}
I_{\rm{int,H}\alpha} = I_{\rm{obs,H}\alpha}10^{0.4A_{\rm{H}\alpha}} = \frac{10^{41.27}}{10^{43.17}} \times I_{\rm{obs,UV}}10^{0.4A_{\rm{UV}}}.
\end{equation} The predicted UV flux map is highly dependent on the assumed relation between $A_{\rm{UV}}$ and $A_{\rm{H}\alpha}$; if we account for extra attenuation towards birth clouds according to \cite{Charlot2000}, the predicted UV flux is slightly higher than observed, and extends towards the South West end of the galaxy. If we use a lower $k_{\rm{H}\alpha}$ based on the continuum $k_{\lambda}$, the flux falls below the noise level of the HST image across the galaxy's spatial extent. \\
\indent Our modelling suggests that it is not possible to predict the observed morphology of the rest-frame UV emission from the combination of the $\rm{H}\alpha$ image and the $A_{\rm{H}\alpha}$ map. This may imply that the recent star-formation (probed by $\rm{H}\alpha$) is attenuated so strongly as to be undetectable in our F606W HST image. In this case, the UV flux that we do observe is tracing star formation on slightly longer timescales. This scenario is consistent with the peak of the stellar mass lying towards the North East of the $\rm{H}\alpha$ flux (see the F140W image). Indeed, qualitatively, we are broadly able to model the observed UV emission by assuming that, before obscuration, the UV light traces the same region as the optical image. If we then apply a dust attenuation map like the one shown in Figure \ref{fig:aha_plot}, we recover a peak of UV emission in the region that is observed. A detailed quantitative comparison of this would require assumptions about the relation between rest-frame UV and rest-frame optical light, which is sensitive to the age and metallicity of the starburst.

\section{SHiZELS-14 in the context of the sub-millimeter galaxy population}\label{sec:smg_comparison}
\begin{figure*}
\includegraphics[scale=0.7]{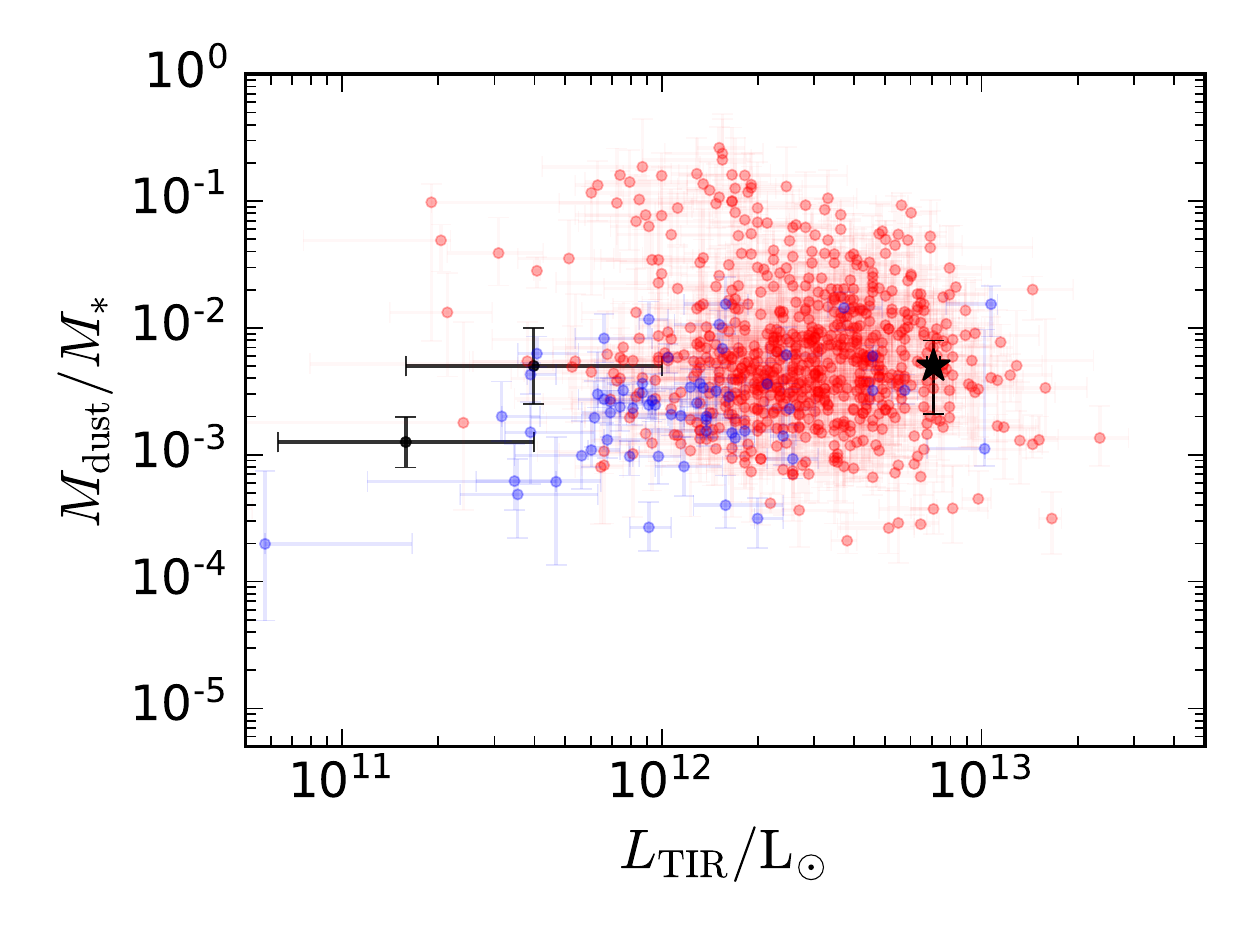}
\includegraphics[scale=0.68]{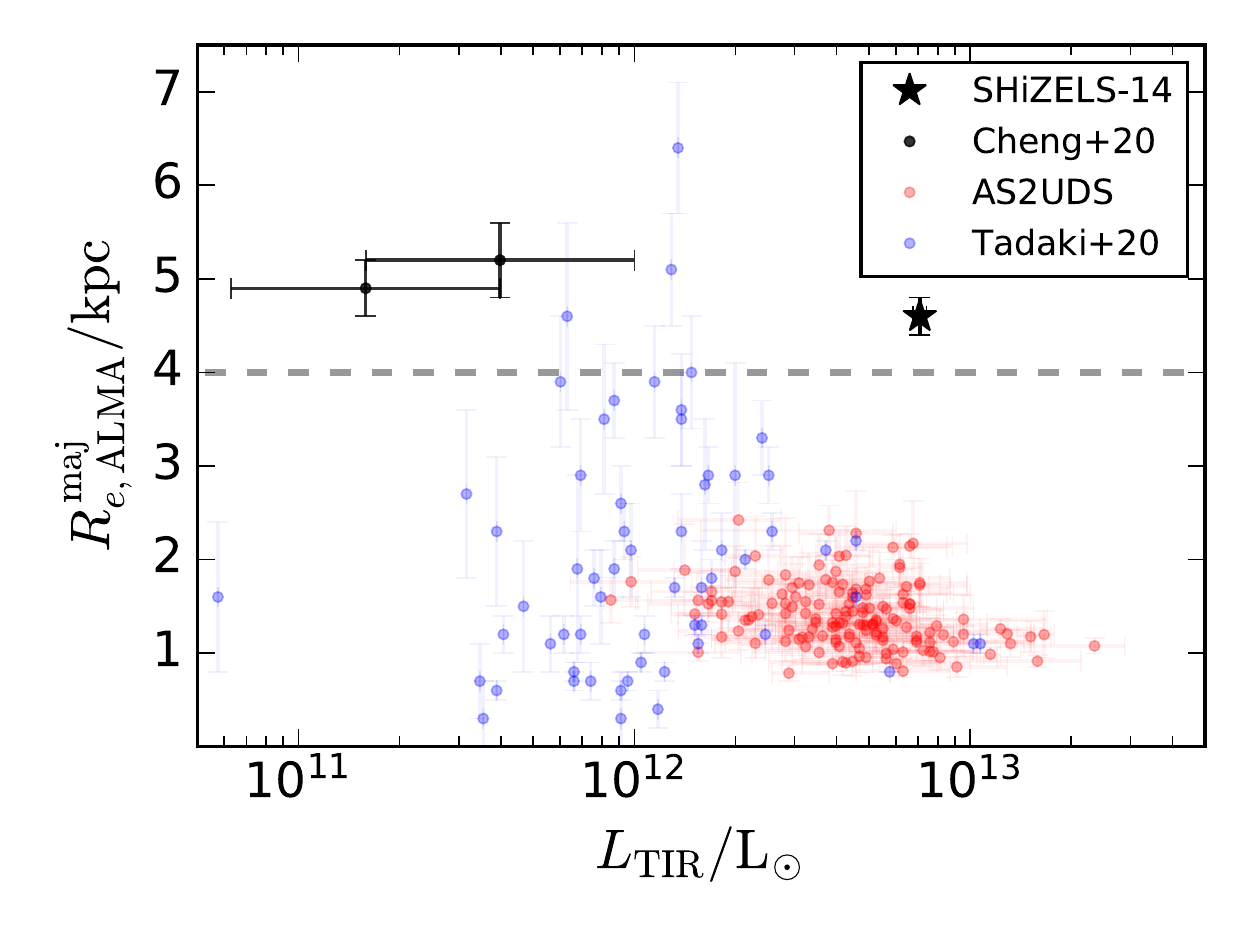}
\caption[]{Left: dust mass-to-stellar mass ratio versus total infrared luminosity, for the AS2UDS sub-millimeter bright galaxies (red circles; \citealt{Dudzeviciute2019}), the subset of stellar mass-selected galaxies from \citep{Tadaki2020a} that are in UDS and have size measurements, and the three TIR-brightest SHiZELS galaxies from \cite{Cheng2020} (black). SHiZELS-14 (black star) is well within the range of both parameters derived for the sub-millimeter galaxy population. Dust mass and stellar mass are derived using {\small{MAGPHYS}} for SHiZELS-14 and the AS2UDS and \cite{Tadaki2020a} samples. Right: effective radius (along the semi-major axis) versus total infrared luminosity for the subsample of these galaxies targeted at higher spatial resolution with ALMA, presented by \cite{Gullberg2019}\footnote{We have corrected a minor error in the table presented by \cite{Gullberg2019}. This error is noted in \cite{Smail2020}. The stated $R_{e}$ values for the case of the fixed $n=1$ fit give the effective radius along the semi-minor axis, rather than the semi-major axis. We correct these using $R_{e,\rm{ALMA}}^{\rm{maj}}=R_{e,\rm{original}}/\rm{axial\,ratio}$.}, with the small statistical correction derived by \cite{Smail2020} applied to the source sizes. The grey line at $R_{e,\,\rm{ALMA}}=4\,\rm{kpc}$ marks the effective radius of the faint extended component that is measured in the stacking analysis of \cite{Gullberg2019}. For both the \cite{Gullberg2019} sample and SHiZELS-14, radii were calculated using two-dimensional S\'{e}rsic fits (with fixed $n=1$) in the image plane. For the remaining SHiZELS galaxies, radii were derived using a curve-of-growth analysis \citep{Cheng2020}. For the \citep{Tadaki2020a} sample, radii were derived using Gaussian model fits in the $uv$-plane. SHiZELS-14 has a much larger effective radius (as measured in the rest-frame FIR) than the majority of the AS2UDS galaxies, though several less FIR-luminous galaxies in the \citep{Tadaki2020a} sample are similarly extended. The extended dust emission suggests that SHiZELS-14 is caught in the mid-stages of a merger.}
\label{fig:smg_plots}
\end{figure*}

SHiZELS-14 was identified by the HiZELS survey, which uses an $\rm{H}\alpha$-based selection and largely probes typical star-forming galaxies, assuming typical extinction. However, it displays a number of extreme properties including high star-formation rate, dust mass and dust attenuation, and a TIR luminosity that places it in the ULIRG regime. In this section, we examine SHiZELS-14 in the context of sub-millimeter galaxies at similar redshifts. \\
\indent We compare SHiZELS-14 to galaxies from the ALMA follow-up of the SCUBA-2 Cosmology Legacy Survey's UKIDSS-UDS field (AS2UDS; \citealt{Stach2018,Stach2019,Dudzeviciute2019}). This is drawn from a $\sim1\,\rm{deg}^{2}$ SCUBA-2 survey. The ALMA pointings target $\sim700$ submillimeter-luminous ($S_{870}\gtrapprox 1\rm{mJy}$) galaxies, with median redshift $z_{\rm{phot}} = 2.61\pm0.09$ \citep{Dudzeviciute2019}. In Figure \ref{fig:smg_plots} (left-hand panel), we show the distribution of dust-to-stellar mass ratio versus total infrared luminosity for the AS2UDS SMGs from the analysis of \cite{Dudzeviciute2019} (red circles). SHiZELS-14 lies above the average of the galaxies in TIR luminosity, but has a fairly unremarkable dust mass-to-stellar mass ratio. In the right-hand panel of the same figure, we show the effective radius of the dust continuum emission along the semi-major axis versus the total infrared luminosity for a subsample of the AS2UDS sources with higher spatial resolution ALMA observations from \cite{Gullberg2019}, with SHiZELS-14 overplotted on the same axes. In the context of these bright SMGs, SHiZELS-14's TIR luminosity is not exceptional. However, it has a dust continuum size that is larger than any of the comparison sample.\\
\indent Like other SMGs, SHiZELS-14 displays a compact core of dust continuum emission. As seen from Figure \ref{fig:four_images}, it also has substantial extended emission, which we are able to resolve due to our deep ALMA imaging. \cite{Gullberg2019} discuss the possibility of an extended component in the AS2UDS SMGs. For sources with SCUBA-2 flux densities brighter than $4\rm{mJy\,beam^{-1}}$ (SHiZELS-14 is in this class), the median flux recovery from the ALMA pointings is $97^{+1}_{-2}$ per cent \citep{Stach2019}. For sources with fainter SCUBA-2 flux densities ($2.5\leq S_{850} \leq2.9\rm{mJy\,beam^{-1}}$), the median flux recovery of those with ALMA detections is $88\pm6$ per cent. These high levels of flux recovery suggest that the rest-frame FIR emission of the AS2UDS SMGs is genuinely very compact, and not dominated by extended emission below the surface brightness limit of the ALMA observations. \cite{Gullberg2019} further characterise the extended emission using a stacking analysis. Their stacked profile is well-fitted by a two-component model, consisting of two S\'ersic profiles of effective radii $\sim0.1''$ and $\sim0.5''$. The extended component accounts for only $13\pm1$ per cent of the integrated modelled emission, on average. While probing to lower surface brightness might increase the effective radii measured (particularly if the fainter flux is substantially more extended than the core), the central $1-2\,\rm{kpc}$ will remain the dominant source of flux. \cite{Smail2020} derive a statistical correction to the AS2UDS source sizes, using a flux-weighted sum of the measured size presented by \cite{Gullberg2019} and a $0.5''$ component with flux density $0.5\,\rm{mJy}$. This increases their source radii by a fairly modest factor of $1.3^{+0.25}_{-0.13}$. We apply this correction to the AS2UDS data, and note that SHiZELS-14 remains an outlier. The $\sim4\,\rm{kpc}$ component that contributes only $\sim10$ per cent of the total flux for the ASUDS SMGs is the dominant component for SHiZELS-14, which displays an extended disk-like structure that is well-fitted by a single two-dimensional S\'ersic profile with $R_{e}^{\rm{maj}}=4.6\,\rm{kpc}$, $n=1$ and $q=0.47$. SHiZELS-14 is genuinely more extended than the majority of the AS2UDS SMGs, perhaps due to being a mid-stage merger.\\
\indent We also compare SHiZELS-14 to a sample of $K$-band-identified, stellar mass-selected ($M_{*}>10^{11}M_{\odot}$), $UVJ$-classified star-forming, intermediate redshift ($z=1.9-2.6$) galaxies studied by \cite{Tadaki2020a}. Unlike the AS2UDS sources, these galaxies were not explicitly selected to be sub-millimeter bright. For the $69$ of their sources that lie in UDS, we make use of the same multi-wavelength parent catalogues used in \cite{Dudzeviciute2019} for the AS2UDS galaxies, and repeat the MAGPHYS SED fitting procedure. We also adopt the effective radii presented by \cite{Tadaki2020a}, obtained via fitting in the $uv$ plane (with fixed $n=1$). Our new MAGPHYS fits and these radii together provide the physical properties of a complementary sample of star-forming galaxies with similar stellar masses and redshifts to SHiZELS-14. \\
\indent In line with their selection, the \cite{Tadaki2020a} sample typically have lower dust masses and TIR luminosities than the AS2UDS sample (median $\log_{10}(L_{\rm{TIR}}/\rm{L}_{\odot})=12.5$ for the AS2UDS sample, compared to $12.0$ for the \cite{Tadaki2020a} sample), as well as lower dust temperatures (median $T_{\rm{dust}}= 40.7\,\rm{K}$ for the AS2UDS sample, compared to  $35.0\,\rm{K}$ for the \cite{Tadaki2020a} sample, though see the caveats noted in \cite{Dudzeviciute2019} regarding discrepancies between MAGPHYS-derived dust temperatures and those inferred from modified black-body fitting). As shown in the right-hand panel of Figure \ref{fig:smg_plots}, the \cite{Tadaki2020a} sample occupies a different region of the $R_{e}$ versus LIR plane to the AS2UDS galaxies. These less TIR-luminous galaxies tend to have larger sizes (median $R_{\rm{eff,ALMA}}^{\rm{maj}}(n=1)=1.8\,\rm{kpc}$, and 11 galaxies have $R_{\rm{eff,ALMA}}^{\rm{maj}}(n=1)>3\,\rm{kpc}$). This result is broadly consistent with the large sizes ($R_{\rm{eff,ALMA}}^{\rm{maj}}\sim5\,\rm{kpc}$) of the broader SHiZELS sample, presented in \cite{Cheng2020}; all but one of these (SHiZELS-14) have lower TIR luminosities than the AS2UDS sample ($\log_{10}(L_{\rm{TIR}}/\rm{L}_{\odot})<12$). The most luminous of the \cite{Tadaki2020a} sample ($\log_{10}(L_{\rm{TIR}}/\rm{L}_{\odot})\gtrapprox12.5$) are just as compact as the AS2UDS sources.\\
\indent Here, we relate the observed morphology of the dust continuum emission to the physical processes taking place within star-forming galaxies around the peak of cosmic star formation. As discussed by \cite{Cheng2020}, the extended dust continuum emission observed in the less-FIR luminous SHiZELS galaxies suggests a dominant component of extended, disk-wide star formation; in contrast, the emission from sub-millimeter selected galaxies appears to be dominated by a compact, nuclear starburst. SHiZELS-14 is an outlier in the sense that it has both a sub-millimeter bright compact core and very extended emission. \cite{Tadaki2020a} show that the most compact galaxies in their sample tend to have high gas fractions (derived via $S_{870\mu\rm{m}}$), and argue that this reflects efficient radial gas inflows. Numerical simulations have long shown that galaxy mergers are capable of triggering tidally-driven gas inflows \citep[][]{Hernquist1989,Barnes1991}, which can cause strong nuclear starbursts \citep[e.g.][]{Mihos1994,Mihos1996,Hopkins2013c,Moreno2015}. However, observations of local galaxies such as the Antennae system demonstrate that galaxy interactions can also trigger widespread star formation that is not limited to a compact, nuclear region \citep{Wang2004}. More recent, high resolution simulations show that these observations can be explained via merger-driven injections of turbulence into the ISM: extended compression results in fragmentation into dense, star-forming gas, and spatially extended starburst activity \citep{Renaud2014,Renaud2015}. \cite{Renaud2015} argue that this process is particularly important in the early and mid stages of a galaxy merger: during the first two simulated pericenter passages, star clusters form kiloparsecs from the galactic nucleus, with the central starburst dominating only from the beginning of the final coalescence. This progression of star formation from extended to compact as the merger unfolds is also consistent with observations of local galaxies \citep{Pan2019}. The extended star formation observed in SHiZELS-14 may therefore suggest that we are viewing the short-lived mid-stages of a merger; this would be consistent with its complex, irregular morphology and dispersion-dominated $\rm{H}\alpha$ kinematics. The similarly TIR-luminous but more compact sources within the AS2UDS samples may comprise galaxies experiencing a wider range of evolutionary stages, including some later-stage mergers.

\section{Conclusions}\label{sec:conclusions}
In this paper, we have presented a study of SHiZELS-14, a $z=2.24$ galaxy originally identified by HiZELS via its $\rm{H}\alpha$ emission. SHiZELS-14 was one of the galaxies selected for high spatial resolution follow-up, due to its proximity to a guide star (for adaptive optics observations), rather than any special properties. However, this galaxy has some intriguing features when resolved at high spatial resolution, particularly at long wavelengths. \\
\indent The global properties of SHiZELS-14 show that it is highly star-forming. SED fits to photometric data indicate a strong burst of star formation within $\sim200$\,Myr of $z=2.24$ and a stellar mass of $10^{11.2\pm0.1}\,\rm{M_{\odot}}$. Fitting the dust SED with modified black body models yields a dust mass of $\rm{M_{\rm{dust}}}=10^{9.0\pm0.1}\,\rm{M_{\odot}}$ and a TIR luminosity of $\log_{10}(L_{\rm{TIR}}/\rm{L_{\odot}})= 12.81\pm0.02$. This bright IR emission places it in the category of a ULIRG, while its strong submillimeter detection shows it is an SMG. SHiZELS-14 lies on the $z\sim2$ IR-radio relation expected for a star-forming galaxy and our extensive multi-wavelength data presents no evidence of AGN activity. \\
\indent FUV, $\rm{H}\alpha$, FIR and radio continuum emission are all used to infer SFR, individually and in combination. We investigate the agreement of widely-used SFR calibrations, globally and in a spatially-resolved manner. Without any dust corrections, the SFRs inferred from FUV and $\rm{H}\alpha$ are $13\pm1\,
\rm{M_{\odot}}\rm{yr}^{-1}$ and $33\pm2\,\rm{M_{\odot}}\rm{yr}^{-1}$, respectively. The SFR inferred from the TIR emission is $950\pm50\,\rm{M_{\odot}}\rm{yr}^{-1}$, and the radio-derived SFR is also in the region $\sim 1000\,\rm{M_{\odot}}\rm{yr}^{-1}$. Thus, SFR inferred from short wavelength light is orders of magnitude lower than that measured at longer wavelengths. This suggests that SHiZELS-14 is affected by a large degree of dust attenuation, in line with its substantial dust mass and FIR flux, and it shares many properties with the known population of high redshift SMGs. \\
\indent We present kpc-scale imaging in the rest-frame FUV and optical (from {\it{HST}}), at FIR-wavelengths (from ALMA), of the $\rm{H}\alpha$ emission line (from SINFONI, on the VLT), and of the radio continuum (from the JVLA). The range of wavelengths probed enables us to detect both unattenuated and dust-reprocessed emission. SHiZELS-14 shows striking, extended emission in both $\rm{H}\alpha$ and the FIR, with $\rm{H}\alpha$-derived effective radius $4.6\pm0.4\,\rm{kpc}$ and a FIR-derived effective radius along the semi-major axis $4.6\pm0.2\,\rm{kpc}$ (axial ratio $q=0.47$). Unlike many SMGs studied at similar redshifts which display compact $1-2\,\rm{kpc}$ cores, SHiZELS-14 displays an extended disk structure in the rest-frame FIR. Our deep imaging enables us to recover directly the fainter emission across extended regions of star formation, which are also traced by $\rm{H}\alpha$. The irregular, extended structures and disordered $\rm{H}\alpha$ kinematics, together with the intense burst of dusty star formation observed, likely reflects ongoing (at $=2.24$) merger activity. \\
\indent The high spatial resolution of our data enables us to study emission on kpc scales, and compare SFRs in a spatially-resolved manner. We show that the SFR surface density maps derived from UV, $\rm{H}\alpha$ and TIR are discrepant across the the extent of the galaxy. Comparison of the $\rm{H}\alpha$ and TIR maps enables us to map the dust attenuation, under the assumption of minimal gradients in dust temperature and optical depth. We find high levels of dust attenuation across the galaxy, with $A_{\rm{H}\alpha}\sim2-3$ in the outskirts, rising to $A_{\rm{H}\alpha}>5$ in the central region. This work highlights the importance of studying galaxies at multiple wavelengths and demonstrates the biases that can be introduced by assuming that calibrations derived using samples of relatively dust-poor galaxies will be appropriate for extremely dusty systems.

\section*{Acknowledgements}
We thank the anonymous reviewer for helpful suggestions that improved the paper. RKC acknowledges funding from an STFC studentship, the Institute for Astronomy, University of Edinburgh, and the John Harvard Distinguished Science Fellowship. PNB is grateful for support from STFC via grant ST/R000972/1. AMS and IRS acknowledge support from STFC (ST/T000244/1). EI acknowledges partial support from FONDECYT through grant N$^\circ$1171710. This work was supported by the National Science Foundation of China (11721303, 11991052) and the National Key R\&D Program of China (2016YFA0400702). CC is supported by the National Natural Science Foundation of China, No. 11803044, 11933003. This work is sponsored (in part) by the Chinese Academy of Sciences (CAS), through a grant to the CAS South America Center for Astronomy (CASSACA). DS acknowledges financial support from Lancaster University through an Early Career Internal Grant A100679. RKC thanks Adam Carnall for assistance with the {\small{BAGPIPES}} SED fitting tool and Richard Bower for helpful discussions during the PhD viva. We thank Jiasheng Huang for providing us with the CLAUDS image of SHiZELS-14. \\
\indent This research is based on observations made with the NASA/ESA {\it{Hubble Space Telescope}} obtained from the Space Telescope Science Institute, which is operated by the Association of Universities for Research in Astronomy, Inc., under NASA contract NAS 5-26555. These observations are associated with program 14919. This paper makes use of the following ALMA data: ADS/JAO.ALMA2015.1.00026.S. ALMA is a partnership of ESO (representing its member states), NSF (USA) and NINS (Japan), together with NRC (Canada), MOST and ASIAA (Taiwan), and KASI (Republic of Korea), in cooperation with the Republic of Chile. The Joint ALMA Observatory is operated by ESO, AUI/NRAO and NAOJ. This paper uses data from SINFONI, based on observations collected at the European Organisation for Astronomical Research in the Southern Hemisphere under ESO programme 084.B-0300. This work is based on data products from observations made with ESO Telescopes at the La Silla Paranal Observatory under ESO programme 179.A-2005 and on data products produced by TERAPIX and the Cambridge Astronomy Survey Unit on behalf of the UltraVISTA consortium.\\
\indent The HSC collaboration includes the astronomical communities of Japan and Taiwan, and Princeton University. The HSC instrumentation and software were developed by the National Astronomical Observatory of Japan (NAOJ), the Kavli Institute for the Physics and Mathematics of the Universe (Kavli IPMU), the University of Tokyo, the High Energy Accelerator Research Organization (KEK), ASIAA, and Princeton University. Funding was contributed by the FIRST program from Japanese Cabinet Office, the Ministry of Education, Culture, Sports, Science and Technology (MEXT), the Japan Society for the Promotion of Science (JSPS), Japan Science and Technology Agency (JST), the Toray Science Foundation, NAOJ, Kavli IPMU, KEK, ASIAA,
and Princeton University.\\
\indent This work is based on observations obtained with MegaPrime/ MegaCam, a joint project of CFHT and CEA/DAPNIA, at the CFHT which is operated by the National Research Council (NRC) of Canada, the Institut National des Science de l’Univers of the Centre National de la Recherche Scientifique (CNRS) of France, and the University of Hawaii. This research uses data obtained through the Telescope Access Program (TAP), which has been funded by the National Astronomical Observatories, Chinese Academy of Sciences, and the Special Fund for Astronomy from the Ministry of Finance. This work uses data products from TERAPIX and the Canadian Astronomy Data Centre. It was carried out using resources from Compute Canada and Canadian Advanced Network For Astrophysical Research (CAN-FAR) infrastructure. These data were obtained and processed as part of CLAUDS, which is a collaboration between astronomers from Canada, France, and China described in \cite{Sawicki2019}.\\
\indent This research has made use of the VizieR catalogue access tool, CDS, Strasbourg, France. The original description of the VizieR service was published in A\&AS 143, 23. The {\it{Herschel}} Extragalactic Legacy Project (HELP) is a European Commission Research Executive Agency funded project under the SP1-Cooperation, Collaborative project, Small or medium-scale focused research project, FP7-SPACE-2013-1 scheme, Grant Agreement Number 607254.

\section*{Data Availability}The data underlying this article will be shared on reasonable request to the corresponding author.

\bibliographystyle{mnras}
\bibliography{Edinburgh}

\appendix
\section{Deriving a rest-frame FIR size from the ALMA data}\label{sec:appedixb}
\begin{figure}
\centering
\includegraphics[scale=0.45]{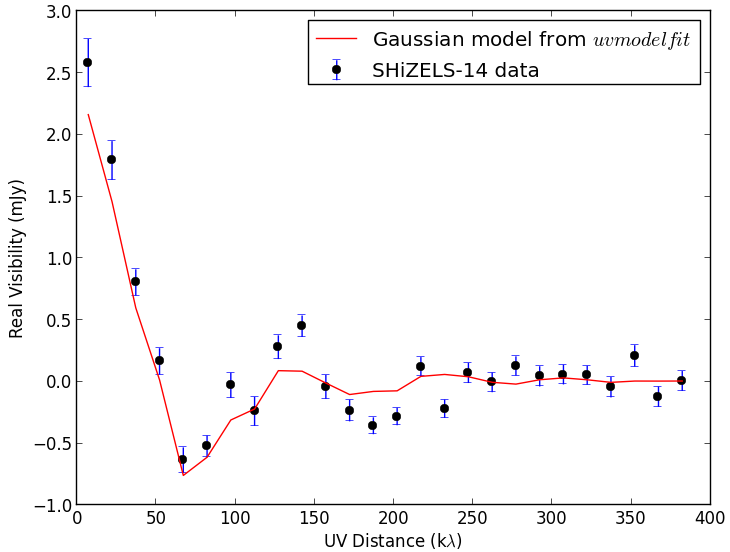}
\caption[]{$uv$-amplitude plot for the ALMA $260\,\rm{GHz}$ data. We fit a Gaussian model with varying axis ratio in the $uv$-plane, using {\small{CASA}}'s {\it{uvmodelfit}} task. The effective radius along the semi-major axis, $R_{e}^{\rm{maj}}$, $4.5\pm0.2\,\rm{kpc}$. Consistent results are obtained using the {\it{uvmultfit}} tool \citep{Marti-Vidal2014}. We also fit an exponential profile and obtain a best-fitting scale length $r_{0}=4.1\pm0.2\,\rm{kpc}$.}
\label{fig:uv_ampl_plot}
\end{figure}

In Figure \ref{fig:uv_ampl_plot} we show a Gaussian model fit to the $260\,\rm{GHz}$ ALMA visibility data. The derived effective radius along the semi-major axis, $R_{e}^{\rm{maj}}=4.5\pm0.2\,\rm{kpc}$, is larger than is typical for SMGs, as shown in Figure \ref{fig:smg_plots}, but broadly in agreement with the effective radius derived from the $\rm{H}\alpha$ image ($4.6\pm0.4\,\rm{kpc}$; \citealt{Swinbank2012a}). We also fit an exponential profile ($f(r)\propto e^{-r/r_{0}}$) with the {\it{uvmultfit}} tool \citep{Marti-Vidal2014} and obtain a best-fitting scale length $r_{0}=4.1\pm0.2\,\rm{kpc}$.
\section{Detailed description of SED fitting with BAGPIPES}\label{sec:appedixa}
\begin{figure}
\centering
\includegraphics[scale=0.34]{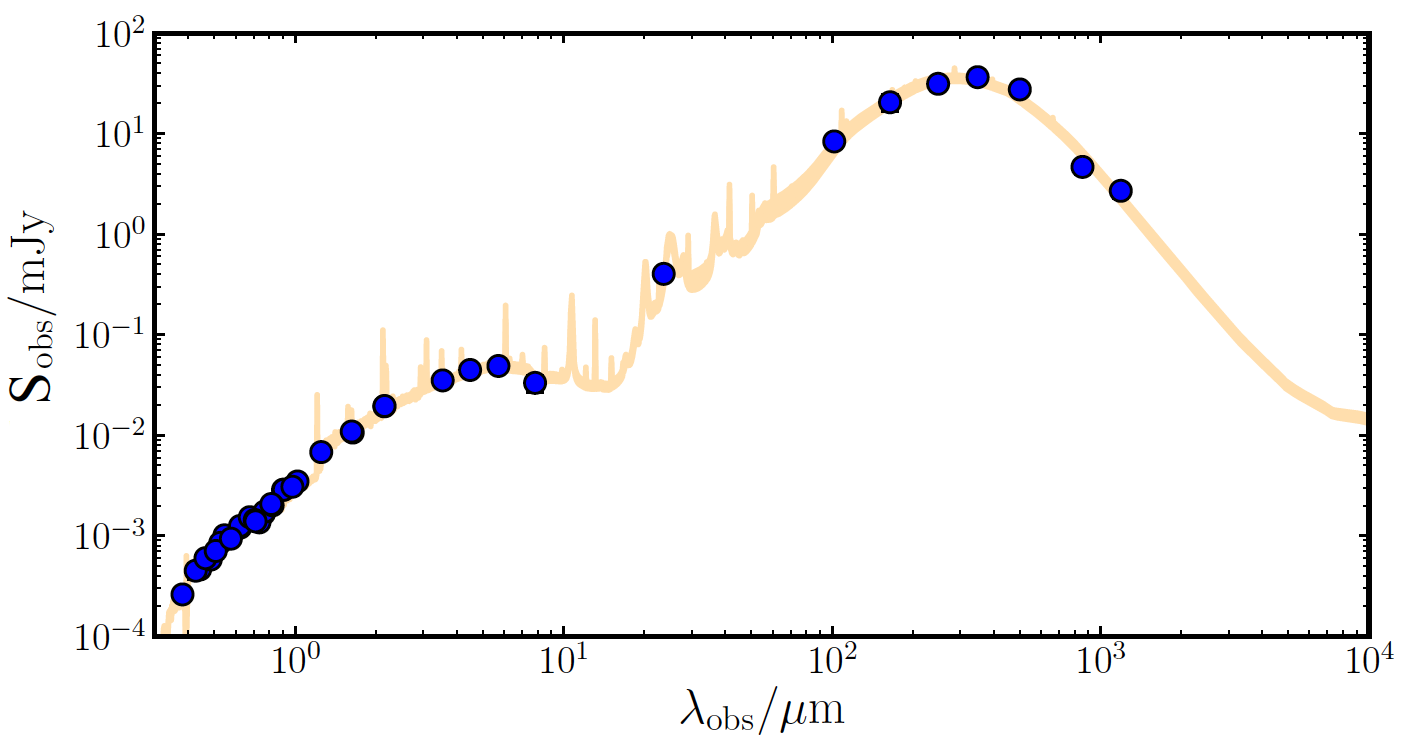}
\caption[SHiZELS-14 SED fit using BAGPIPES]{Data presented in Table \ref{Table:hizels_laigle}, fitted with the {\small{BAGPIPES}} code \citep{Carnall2018}, using a double power law star formation history. Error bars are plotted on the data points, but are small. The fitting yields $\rm{SFR}=660\pm60\,M_{\odot}/\rm{yr}$ and $\log_{10}M_{*}/M_{\odot}=11.1\pm0.1$, in good agreement with the estimates from {\small{MAGPHYS}}.}
\label{fig:bagpipes_sed2}
\end{figure}
To assess the sensitivity of our derived physical parameters to our choice of SED fitting code, we refit the photometry with another code. Our fitting makes use of the 2016 version of the BC03 SSP templates, with a \cite{Kroupa2002} IMF (note that the difference between a Kroupa and Chabrier IMF is negligible). Nebular emission is computed using the {\small{CLOUDY}} photoionization code \citep{Ferland2017}, following \cite{Byler2017}. {\small{CLOUDY}} is run using each SSP template as the input spectrum. Dust grains are included using {\small{CLOUDY}}'s `ISM' prescription, which implements a grain-size distribution and abundance pattern that reproduces the observed extinction properties for the ISM of the Milky Way. We select a \cite{Calzetti2000} dust attenuation curve. Dust emission includes both a hot dust component from HII regions and a grey body component from the cold, diffuse dust. \\
\indent We impose a wide dust attenuation prior, $A_{v}=[0,6]$, which gives the code the option to fit a high degree of attenuation. Following \cite{Draine2007}, we fit three parameters that affect the shape of the dust SED: $U_{\rm{min}}$, the lower limit of the starlight intensity; $\gamma$, the fraction of stars at $U_{\rm{min}}$; and $q_{\rm{PAH}}$, the mass fraction of polycyclic aromatic hydrocarbons. Our priors on these parameters are broad, to allow the model the option to fit a hot dusty galaxy: $U_{\rm{min}}=[0,25]$, $\gamma=[0,1]$, and $q_{\rm{PAH}}=[0,10]$. We also fit $\eta$, the multiplicative factor on $A_{V}$ for stars in birth clouds, using the range $\eta=[1,5]$. We allow metallicity to vary in the range $Z=[0,2.5]Z_{\odot,\rm{old}}$, where $Z_{\odot,\rm{old}}$ denotes solar models prior to \cite{Asplund2009}. We fix the redshift at $z=2.2418$, since this is known from the SINFONI spectrum. \\
\indent We experiment with various star-formation history (SFH) parametrisations, which yield very similar fits to the spectrum and consistent values for stellar mass, $\log_{10} M_{*}/\rm{M_{\odot}}=11.1\pm0.1$. All parametrisations, even those allowing multiple bursts, favour a recent (at $z=2.24$), rapid burst of star formation in which the vast majority of the stellar mass is formed. In Figure \ref{fig:bagpipes_sed2}, we plot a representative fit to the photometry. This particular model uses a double power law SFH parametrisation. The posterior estimate for the star-formation rate is $\rm{SFR}=660\pm60\,\rm{M_{\odot}}\rm{yr}^{-1}$, and the estimated specific star-formation rate (sSFR) is $\log_{10}(\rm{sSFR}/yr^{-1})=-8.25\pm0.11$. Note that the SFR is more sensitive than the stellar mass to the parametrisation of the SFH and the data included in the fit, and averaging over multiple SFH models increases the uncertainty on the SFR to $\sim100\,\rm{M_{\odot}}\rm{yr}^{-1}$. The posterior estimate for the dust attenuation in the $V$-band is $A_{v}=1.8\pm0.1$. All of these derived physical parameters are consistent with the estimates from {\small{MAGPHYS}}, which indicates that our fitting is robust to choice of SED fitting code.

\begin{table}
\begin{center}
\begin{tabular}{l|c|c}
Instrument/Telescope  & Filter & Measurement \\ 
(Survey)  &   & ($\mu\rm{Jy}$) \\
\hline
MegaCam/CFHT & $u^{*}$ & $0.26\pm0.04$ \\
\hline
Suprime-Cam/Subaru & $B$ & $0.46\pm0.03$ \\
& $V$ & $1.01\pm0.05$ \\
& $r$ & $1.21\pm0.05$ \\
& $i_{\rm{+}}$ & $1.66\pm0.05$ \\
& $z_{\rm{+}}$ & $2.87\pm0.16$ \\
& $z_{\rm{++}}$ & $2.87\pm0.07$ \\
& $\rm{IA}427$ & $0.45\pm0.08$ \\
& $\rm{IA}464$ & $0.60\pm0.09$ \\
& $\rm{IA}484$ & $0.58\pm0.08$ \\
& $\rm{IA}505$ & $0.71\pm0.09$ \\
& $\rm{IA}527$ & $0.85\pm0.06$ \\
& $\rm{IA}574$ & $0.94\pm0.10$ \\
& $\rm{IA}624$ & $1.25\pm0.08$ \\
& $\rm{IA}679$ & $1.53\pm0.13$ \\
& $\rm{IA}709$ & $1.44\pm0.09$ \\
& $\rm{IA}738$ & $1.36\pm0.10$ \\
& $\rm{IA}767$ & $1.74\pm0.12$ \\
& $\rm{IA}827$ & $2.01\pm0.14$ \\
& $\rm{NB}711$ & $1.39\pm0.16$ \\
& $\rm{NB}816$ & $2.07\pm0.15$ \\
\hline
HSC/Subaru & $Y_{\rm{HSC}}$ & $3.1\pm0.2$ \\
\hline
VIRCAM/VISTA &  $Y$ & $3.47\pm0.07$ \\
(UltraVISTA-DR2) & $J$ & $6.80\pm0.10$ \\
& $H$ & $10.64\pm0.16$ \\
& $K_{s}$ & $19.40\pm0.14$ \\ 
\hline
WIRCam/CFHT & $K_{\rm{sw}}$ & $19.6\pm0.8$ \\
& $H_{\rm{w}}$ & $10.9\pm0.7$ \\
\hline
{\it{Spitzer}}/IRAC & $3.6\,\mu \rm{m}$ & $35.1\pm0.3$ \\
(SPLASH) & $4.5\,\mu \rm{m}$ & $44.6\pm0.3$ \\
& $5.8\,\mu \rm{m}$ & $48.9\pm3.6$ \\
& $8\,\mu \rm{m}$ & $33.2\pm6.1$ \\
\hline
{\it{Spitzer}}/MIPS & $24\,\mu \rm{m}$ & $403\pm17$ \\
\hline
Herschel-HerMES/ Oliver+12& $100\,\mu \rm{m}$ & $8.4\pm0.9\,(\rm{mJy})$ \\
HELP catalogue values & $160\,\mu \rm{m}$ & $20.5\pm3.7\,(\rm{mJy})$ \\
  & $250\,\mu \rm{m}$ & $31.3\pm2.2\,(\rm{mJy})$ \\
 & $350\,\mu \rm{m}$ & $36.5\pm2.5\,(\rm{mJy})$ \\
 & $500\,\mu \rm{m}$ & $27.5\pm2.7\,(\rm{mJy})$ \\
\hline
ALMA Band 6, this paper & $260\,\rm{GHz}$ & $2.7\pm{0.2}\,(\rm{mJy})$ \\
ALMA Band 7, Scoville+14 & $350\,\rm{GHz}$ & $4.7\pm{0.8}\,(\rm{mJy})$ \\
SCUBA-2, Simpson+19 & $350\,\rm{GHz}$ & $5.4\pm{1.3}\,(\rm{mJy})$ \\
\hline
JVLA, This paper & $6\,\rm{GHz}$ & $20\pm2$ \\
JVLA, Smol\v ci\'c+17 & $3\,\rm{GHz}$ & $68\pm4$ \\
VLA, Schinnerer+10 & $1.4\,\rm{GHz}$ & $122\pm13$ \\
\end{tabular}
\caption[]{Compilation of existing and new measurements of SHiZELS-14, with source. Unless otherwise stated, the data are taken these from the tables of \cite{Laigle2016}, adopting their values calculated within a $3''$ diameter aperture.}
\label{Table:hizels_laigle}
\end{center}
\end{table}

\bsp	
\label{lastpage}
\end{document}